\documentclass[preprint,twocolumn]{aastex6}
\shorttitle{MILDLY RELATIVISTIC EJECTA}
\shortauthors{SUZUKI, MAEDA, \& SHIGEYAMA}
\begin{document}
\title{HYDRODYNAMICAL INTERACTION OF MILDLY RELATIVISTIC EJECTA WITH AN AMBIENT MEDIUM}
\author{AKIHIRO SUZUKI\altaffilmark{1}, KEIICHI MAEDA\altaffilmark{1,2}, and TOSHIKAZU SHIGEYAMA\altaffilmark{3}}
\altaffiltext{1}{Department of Astronomy, Kyoto University, Kitashirakawa-Oiwake-cho, Sakyo-ku, Kyoto, 606-8502, Japan.}
\altaffiltext{2}{Kavli Institute for the Physics and Mathematics of the Universe (WPI), Todai Institutes for Advanced Study, 
The University of Tokyo, 5-1-5 Kashiwanoha, Kashiwa, Chiba 277-8583, Japan.}
\altaffiltext{3}{Research Center for the Early Universe, School of Science, University of Tokyo, Bunkyo-ku, Tokyo, 113-0033, Japan.}
\begin{abstract}
Hydrodynamical interaction of spherical ejecta freely expanding at mildly relativistic speeds into an ambient cold medium is studied in semi-analytical and numerical ways to investigate how ejecta produced in energetic stellar explosions dissipate their kinetic energy through the interaction with the surrounding medium. 
We especially focus on the case in which the circumstellar medium is well represented by a steady wind at a constant mass-loss rate having been ejected from the stellar surface prior to the explosion. 
As a result of the hydrodynamical interaction, the ejecta and circumstellar medium are swept by the reverse and forward shocks, leading to the formation of a geometrically thin shell. 
We present a semi-analytical model describing the dynamical evolution of the shell and compare the results with numerical simulations. 
The shell can give rise to bright emission as it gradually becomes transparent to photons. while it is optically thick. 
We develop an emission model for the expected emission from the optically thick shell, in which photons in the shell gradually diffuse out to the interstellar space. 
Then, we investigate the possibility that radiation powered by the hydrodynamical interaction is the origin of an underluminous class of gamma-ray bursts. 
\end{abstract}

\section{INTRODUCTION}\label{sec:intro}
Collisions of two cold media and subsequent formation of shocks propagating into both sides of the contact surface provide an efficient way to convert the kinetic energy of the flows into the internal energy of the shocked media. 
This process is thought to be a plausible source of high-energy emission in various astrophysical situations.  
Stellar explosions are one such astrophysical site. 
At the final evolutionary stage of massive stars, the collapse of the iron core liberates its gravitational energy, leading to the violent explosion of the star. 
The stellar mantle ejected as a result of the explosion could give rise to bright emission, which is observed as a core-collapse supernova (CCSN). 
A small fraction of CCSNe is known to be accompanied by an intense burst of gamma-ray photons, which is called a long-duration gamma-ray burst (long GRB) \citep[see,][for reviews]{1999PhR...314..575P,2006RPPh...69.2259M,2006ARA&A..44..507W,2012grbu.book..169H,2015PhR...561....1K}. 
A great deal of attention has been paid to the physical link between CCSNe and GRBs. 

In these phenomena, the erupted materials eventually collide with the surrounding gas, i.e., the circumstellar medium (CSM). 
The hydrodynamical interaction of the ejecta and the CSM is of vital importance in understanding electromagnetic signals from some kinds of SNe. 
The shocked SN ejecta and CSM form a shell at the interface between these two media \citep{1981ApJ...251..259C}. 
For a dilute CSM, the shell would be optically thin and thus could give rise to non-thermal radiation from highly energetic electrons, which are accelerated at the shock front, via synchrotron and/or inverse Compton processes. 
Non-thermal emission from SNe has been extensively investigated in many theoretical works  \citep[e.g.,][]{1982ApJ...259..302C,1998ApJ...499..810C,2006ApJ...651..381C}. 
Radio and X-ray observations of several SNe have been carried out and succeeded in constraining the CSM density by using those theoretical models \citep[e.g.,][]{1989ApJ...336..421W,1990ApJ...364..611W,1994ApJ...432L.115V,2002ARA&A..40..387W,2002ApJ...572..932P,2006ApJ...651.1005S,2012ApJ...758...81M}. 

When the CSM is sufficiently dense, the shell would become optically thick and serve as a source of bright thermal emission in early stages of its dynamical evolution. 
In fact, bright emission found in a special class of SNe with prominent hydrogen narrow lines in their spectra, type IIn SNe \citep[see, e.g.,][for a review]{1997ARA&A..35..309F,2014ARA&A..52..487S}, is explained by the hydrodynamical interaction. 
Moreover, there is increasing evidence that some events spectroscopically classified as normal SNe must have exploded in a dense CSM or with an extended envelope attached \citep{2013Natur.494...65O,2014Natur.509..471G,2014ApJ...789..104O}, suggesting the important roles played by the ejecta-CSM interaction especially at the very beginning of the dynamical evolution of SN ejecta. 
It is also claimed that the first electromagnetic signal from an SN explosion, i.e., the shock breakout \citep{1974ApJ...187..333C,1978ApJ...223L.109K,1978ApJ...225L.133F}, can be significantly prolonged and become brighter in the presence of a dense CSM \citep[e.g.,][]{1973ApJ...180L..65F,1977ApJS...33..515F,2011ApJ...729L...6C}. 
There are an increasing number of theoretical studies aiming at establishing appropriate light curve models of the early emission from SNe having exploded in a dense CSM \citep[e.g.,][]{2011MNRAS.414.1715B,2011MNRAS.415..199M,2012ApJ...757..178G,2012ApJ...759..108S,2014ApJ...780...18G,2014ApJ...788..113S,2015A&A...575L..10M}. 

In the context of long GRBs, the CSM interaction might produce some underluminous events. 
A significant fraction of GRBs with spectroscopically identified SN components shows isotropic gamma-ray luminosities 3-5 orders of magnitude lower, $L_\gamma=10^{46-48}$ erg s$^{-1}$, than the typical luminosity of long GRBs, therefore making these events classified as low-luminosity GRBs (LLGRBs). 
Well-known examples include GRB 980425/SN 1998bw \citep{1998Natur.395..663K,1998Natur.395..670G}, GRB 060218/SN 2006aj \citep{2006Natur.442.1008C,2006Natur.442.1011P,2006Natur.442.1014S,2006Natur.442.1018M}, and GRB 100316D/SN 2010bh \citep{2011MNRAS.411.2792S}. 
Several theoretical models have been proposed to explain these nearby peculiar events. 
While normal GRBs at cosmological distances require an ultra-relativistic jet emanating from a massive star, a less-collimated outflow, a chocked jet, or even spherical ejecta moving at mildly relativistic speeds could account for the gamma-ray luminosities of LLGRBs \citep[see, e.g.,][]{2007MNRAS.375..240L,2007ApJ...659.1420T,2007ApJ...664.1026W,2007ApJ...667..351W}. 
Follow-up observations suggest that a mildly relativistic blast wave with shock Lorentz factors of a few is likely to produce their radio afterglows \citep{1998Natur.395..663K,2004Natur.430..648S,2006Natur.442.1014S}, making the relativistic shock breakout model, in which a blast wave driven by relativistic ejecta emerges from the photosphere in a dense CSM, the most popular scenario \citep{1998Natur.395..663K,2001ApJ...551..946T,2006Natur.442.1008C,2007MNRAS.375..240L,2007ApJ...667..351W,2012ApJ...747...88N}. 

How the relativistic ejecta responsible for the gamma-ray emission are produced and how much kinetic energy is distributed in different layers of the ejecta are still debated. 
Self-similar solutions for spherical flows have played important roles in determining the density profile and the kinetic energy distribution of the ejecta. 
\cite{s60} found a series of self-similar solutions describing a strong shock wave propagating in a stellar atmosphere with a power-law density profile. 
\cite{1999ApJ...510..379M} obtained the density structure of SN ejecta resulting from stellar mantle after the passage of a blast wave based on the Sakurai's self-similar solutions. 
\cite{2001ApJ...551..946T} extended the work to study the properties of trans-relativistic ejecta produced in highly energetic SNe. 
\cite{2005ApJ...627..310N} discovered the ultra-relativistic counterpart of the Sakurai's self-similar solutions and discussed the kinetic energy distribution of the ejecta. 

The ejecta could also be produced as a result of aspherical energy deposition or penetration of a jet. 
\cite{2011ApJ...739L..55B,2011ApJ...740..100B} investigated the time required for a jet injected at the core of a massive star to penetrate the star. 
They found that jets with energy injection rates inferred from gamma-ray observations of LLGRBs could not penetrate the progenitor star and concluded that the mechanism to produce LLGRBs seems to be different from normal long GRBs. 
The failed jet hypothesis has been examined by \cite{2012ApJ...750...68L}, who carried out 2D hydrodynamic simulations of the jet propagation in a massive star with several values of the energy injection rate. 
They confirmed that a failed jet can produce relativistic ejecta with steep kinetic energy distributions. 
Relativistic ejecta could also be realized as a cocoon component associated with successful penetration of a massive star by an ultra-relativistic jet \citep{2002MNRAS.337.1349R,2005ApJ...629..903L}. 
By carrying out 2D hydrodynamic simulations, \cite{2013ApJ...764L..12S} demonstrated that the hydrodynamical interaction between the cocoon expanding at mildly relativistic speeds and a dense CSM could produce bright X-ray emission similar to GRB 060218. 
\cite{2015ApJ...807..172N} recently argued that LLGRBs were produced when an extremely dense stellar envelope surrounding the progenitor star could stop the propagation of an ultra-relativistic jet emanating from the core. 

From an observational point of view, follow-up radio observations are a key to probing the kinetic energy distribution of SN ejecta at later epochs \citep[e.g.,][]{1998Natur.395..663K,2004Natur.430..648S,2006Natur.442.1014S,2013ApJ...778...18M,2014ApJ...797..107M}. 
These studies estimated the kinetic energy of the blast wave responsible for radio synchrotron emission, which can be combined with the kinetic energy of SN ejecta emitting optical photons to obtain the distribution of the kinetic energy in the velocity space. 
The kinetic energy of the radio emitting blast wave is much smaller than that of SN ejecta for normal striped-envelope SNe, making the kinetic energy distribution very steep.
LLGRBs exhibit shallower kinetic energy distributions than normal SNe, suggesting that the presence of some sort of central engine activity distinguishes them from normal SNe.  
However, their kinetic energy distributions clearly look different from almost flat distributions inferred for normal GRBs. 
The authors conclude that LLGRBs are driven by a distinct central engine from normal GRBs. 

Investigations of the dynamical evolution of freely expanding ejecta with a wide variety of the kinetic energy distribution colliding with an ambient medium are therefore of crucial importance in examining its potential to produce bright X-ray and gamma-ray emission as well as radio synchrotron emission. 
Self-similar solutions have been used in describing the shocks forming as a result of the hydrodynamical interaction between the ejecta and the ambient medium. 
\cite{1981ApJ...251..259C} discovered a series of self-similar solutions describing the hydrodynamical interaction in the non-relativistic regime. 
The ultra-relativistic extension of this study was done by \cite{2006ApJ...645..431N}. 
However, the intermediate regime, in which the ejecta and the shocks are moving at mildly relativistic speeds, has received little attention despite its recently increasing interest. 

In this paper, we consider spherical ejecta freely expanding at mildly relativistic speeds and investigate the hydrodynamical interaction with an ambient medium in semi-analytical and numerical ways. 
In Section \ref{sec:evolution}, the dynamical evolution of the gas is considered in a semi-analytical way. 
Numerical simulations are performed and compared with the semi-analytical model. 
The results are presented in Section \ref{sec:numerical}. 
We develop an analytical model of the emission powered by the interaction, and the calculated light curves are compared with the X-ray light curve of GRB 060218 in Section \ref{sec:emission}. 
Finally, we conclude the paper in Section \ref{sec:conclusions}. 
In the following, we use the unit $c=1$, where $c$ denotes the speed of light. 

\section{Dynamical evolution of ejecta}\label{sec:evolution}
We consider relativistic flows with spherical symmetry. 
The hydrodynamical variables, the velocity $\beta(t,r)$, density $\rho(t,r)$, and pressure $p(t,r)$, are expressed as functions of time $t$ and the radial coordinate $r$. 
The following equations describe the temporal evolution of these variables,
\begin{equation}
\frac{\partial (\rho\Gamma)}{\partial t}+\frac{1}{r^2}\frac{\partial (r^2\rho\Gamma\beta)}{\partial r}=0,
\label{eq:continuity}
\end{equation}
\begin{equation}
\frac{\partial (\rho h\Gamma^2\beta)}{\partial t}+\frac{1}{r^2}\frac{\partial (r^2\rho h\Gamma^2\beta^2)}{\partial r}+\frac{\partial p}{\partial r}=0,
\label{eq:momentum}
\end{equation}
and
\begin{equation}
\frac{\partial (\rho h \Gamma^2-p)}{\partial t}+\frac{1}{r^2}\frac{\partial (r^2\rho h \Gamma^2\beta)}{\partial r}=0,
\label{eq:energy}
\end{equation}
where the Lorentz factor $\Gamma$ is defined as follows, 
\begin{equation}
\Gamma=\frac{1}{\sqrt{1-\beta^2}}.
\end{equation}
In this work, the gas is treated as an ideal gas with an adiabatic index of $\gamma=4/3$. 
Thus, the specific enthalpy $h$ is given by
\begin{equation}
h=1+\frac{\gamma}{\gamma-1}\frac{p}{\rho}=1+\frac{4p}{\rho}.
\label{eq:enthalpy}
\end{equation}

\subsection{Initial Condition}
The density profile of freely expanding ejecta can generally be expressed as a function of the 4-velocity $\Gamma\beta$. 
Specifically, we consider density profile described by a broken power-law function with a break at $\Gamma\beta=\Gamma_\mathrm{br}\beta_\mathrm{br}$. 
The low-velocity component of the ejecta obeys a flat distribution, $\propto (\Gamma\beta)^0$, while the high-velocity component obeys a power-law distribution with an exponent $-n$. 
Thus, the density profile is given by the following function,
\begin{equation}
\rho_\mathrm{ej}(t,r)=
\left\{
\begin{array}{ccl}
\rho_0\left(\frac{t}{t_0}\right)^{-3}
\left(\frac{\Gamma_\mathrm{br}\beta_\mathrm{br}}{\Gamma_\mathrm{max}\beta_\mathrm{max}}\right)^{-n}&&\\
\mathrm{for}\ \ \Gamma\beta\leq \Gamma_\mathrm{br}\beta_\mathrm{br}&&,
\\
\rho_0\left(\frac{t}{t_0}\right)^{-3}
\left(\frac{\Gamma\beta}{\Gamma_\mathrm{max}\beta_\mathrm{max}}\right)^{-n}&&\\
\mathrm{for}\ \ \Gamma_\mathrm{br}\beta_\mathrm{br}<\Gamma\beta\leq \Gamma_\mathrm{max}\beta_\mathrm{max}&&.
\end{array}
\right.
\label{eq:ejecta_profile}
\end{equation}
We note that the flat part is introduced so that the total kinetic energy of the ejecta does not diverge. 
Because we are interested in ejecta whose trans-relativistic part obeys a power-law function, we assume a non-relativistic break velocity and set to $\Gamma_\mathrm{br}\beta_\mathrm{br}=0.1$. 
As we will see below, the reverse shock does not reach the layer traveling at this velocity even when the maximum ambient gas density and the minimum ejecta energy are assumed. 
Since the gas is assumed to be freely expanding, the velocity is given by the radius $r$ divided by the elapsed time $t$,
\begin{equation}
\beta(t,r)=\frac{r}{t}.
\end{equation}
Furthermore, the ejecta are assumed to be cold, i.e., the pressure is negligibly small and does not affect the dynamical evolution. 

Some parameters, $\Gamma_\mathrm{max}$, $\beta_\mathrm{max}$, and $n$, characterizing the density profile have been introduced. 
The ejecta start interacting with the surrounding gas at $t=t_0$. 
The maximum Lorentz factor of the ejecta at the initial stage of the dynamical evolution is denoted by $\Gamma_\mathrm{max}$. 
The corresponding maximum velocity $\beta_\mathrm{max}$ is given by
\begin{equation}
\beta_\mathrm{max}=\sqrt{1-\frac{1}{\Gamma_\mathrm{max}^2}}.
\end{equation}
The exponent $n$ describes the distribution of the kinetic energy in the ejecta, i.e., ejecta with a larger value of $n$ have less kinetic energy in the outermost layer. 
The normalization $\rho_0$ of the density profile is determined in the following way. 
We define the kinetic energy $E_\mathrm{rel}$ of the relativistic component of the ejecta ($\Gamma\beta\geq 1$). 
The characteristic density $\rho_0$ is proportional to the kinetic energy $E_\mathrm{rel}$. 
The following integration gives the relation between these two quantities, $\rho_0$ and $E_\mathrm{rel}$, for a given set of $t_0$ and $\Gamma_\mathrm{max}$, 
\begin{equation}
E_\mathrm{rel}=4\pi\int_{\beta_\mathrm{min}t}^{\beta_\mathrm{max}t}\rho_\mathrm{ej}\Gamma(\Gamma-1)r^2dr.
\end{equation}
Here $\beta_\mathrm{min}$ is set to $1/\sqrt{2}$, which gives the 4-velocity of unity, $\Gamma\beta=1$. 
We denote the kinetic energy in units of $10^{51}$ erg by $E_\mathrm{rel,51}$. 
The total kinetic energy of the ejecta obtained by the following integral,
\begin{equation}
E_\mathrm{tot}=4\pi\int_{0}^{\beta_\mathrm{max}t}\rho_\mathrm{ej}\Gamma(\Gamma-1)r^2dr,
\end{equation}
highly depends on the exponent $n$. 
The mass of the trans-relativistic ejecta and the total mass of the ejecta can be calculated in similar ways,
\begin{equation}
M_\mathrm{rel}=4\pi
\int_{\beta_\mathrm{min}t}^{\beta_\mathrm{max}t}\rho_\mathrm{ej}\Gamma r^2dr,
\end{equation}
and
\begin{equation}
M_\mathrm{tot}=4\pi
\int_{0}^{\beta_\mathrm{max}t}\rho_\mathrm{ej}\Gamma r^2dr. 
\end{equation}
These quantities, $E_\mathrm{tot}$, $M_\mathrm{rel}$, and $M_\mathrm{tot}$, are proportional to the kinetic energy $E_\mathrm{rel}$. 

As noted in the previous section, several earlier studies \citep{1999ApJ...510..379M,2001ApJ...551..946T,2005ApJ...627..310N} investigated values of the exponent $n$ describing ejecta realized as a result of shock propagation in a massive star in different regimes, i.e., non-relativistic, trans-relativistic, and ultra-relativistic regimes. 
These studies focus on the emergence of a strong shock wave from a stellar atmosphere with spherical symmetry or in plane-parallel geometry. 
\cite{2001ApJ...551..946T} showed that trans-relativistic and ultra-relativistic parts of the ejecta are characterized by a power-law function of the 4-velocity with an exponent $-5.2$ and $-1.1$.  
For the emergence of an ultra-relativistic shock wave from a stellar atmosphere, the self-similar solution obtained by \cite{2005ApJ...627..310N} shows the same dependence as \cite{2001ApJ...551..946T}. 
On the other hand, quasi-spherical ejecta with different kinetic energy distributions could be produced as a result of the jet propagation or a failed jet in a massive star. 
Therefore, in this work, we consider ejecta with various values of the exponent, $n=1,$ $2$, $3$, $4$, and $5$ and the maximum Lorentz factor, $\Gamma_\mathrm{max}=10,$ $5$ and $3$, to examine how different sets of these parameters affect the kinetic energy dissipation of the ejecta due to the collision with the ambient medium. 
The quantities, $E_\mathrm{tot}$, $M_\mathrm{rel}$, and $M_\mathrm{tot}$ for $E_\mathrm{rel,51}=1$ and different sets of the parameters are summarized in Table \ref{table:mass}. 
The total mass of the ejecta can be very small especially for small values of $n$. 
This is because we focus on ejecta traveling at relatively high velocities, $\Gamma\beta>0.1$ and normalize the distribution by using the relativistic component. 
The total masses in Table \ref{table:mass} do not necessarily reflect the total mass of the whole supernova ejecta. 
It is not our purpose to reproduce the density distribution of supernova ejecta precisely.

The ejecta collide with an ambient medium whose density profile is a power-law function of the radius with an exponent $-k$,
\begin{equation}
\rho_\mathrm{a}(r)=Ar^{-k}.
\label{eq:csm}
\end{equation}
In particular, we regard the ambient medium as a steady wind at a constant mass-loss rate and thus set the exponent $k$ to $k=2$ throughout this work. 
The semi-analytical model introduced below can be applied as long as the forward shock decelerates in the ambient medium, i.e., $k<3$. 
For $k>3$, the forward shock is expected to accelerate and leave behind the interface between the ejecta and the ambient medium. 
The time dependence of physical quantities of the shocked gas can vary depending on the exponent $k(<3)$. 
However, the temporal behaviors would not be qualitatively different.  
For a given set of the mass-loss rate $\dot{M}$ and the wind velocity $v_\mathrm{w}$, the coefficient $A$ is expressed as follows,
\begin{equation}
A=\frac{\dot{M}}{4\pi v_\mathrm{w}}=5.0\times 10^{11}\dot{M}_{-5}v_\mathrm{w,3}^{-1} \ \mathrm{g\ cm}^{-1},
\end{equation}
where $\dot{M}=10^{-5}\dot{M}_{-5}\ M_\odot\ \mathrm{yr}^{-1}$ and $v_\mathrm{w}=10^3\ v_\mathrm{w,3}\ \mathrm{km}\ \mathrm{s}^{-1}$. 
Hereafter, $A_{\star}$ stands for the parameter $A$ in units of $5\times 10^{11}$ g cm$^{-1}$, $A=5\times 10^{11}A_\star\ \mathrm{g}\ \mathrm{cm}^{-1}$ and we use $A_\star$ as the parameter describing the density of the ambient medium. 
The ambient medium is also assumed to be cold. 

When the ejecta start colliding with the ambient medium at $t=t_0$, the ejecta fill a region from $r=0$ to $r=\beta_\mathrm{max}t_0$ and the outermost layer is adjacent to the ambient medium at $r=\beta_\mathrm{max}t_0$. 

\begin{table}
\begin{center}
  \caption{Total mass of relativistic ejecta in units of solar mass for $E_\mathrm{rel,51}=1$}
\begin{tabular}{ccrrrr}
\hline\hline
$\Gamma_\mathrm{max}$&$n$&$E_\mathrm{tot} [10^{51}\mathrm{erg}]$&$M_\mathrm{rel}\ [M_\odot]$&$M_\mathrm{tot}\ [M_\odot]$\\
\hline
10&1&1.118&3.783$\times10^{-4}$&7.618$\times10^{-4}$\\
10&2&1.442&5.470$\times10^{-4}$&2.761$\times10^{-3}$\\
10&3&2.473&6.976$\times10^{-4}$&1.548$\times10^{-2}$\\
10&4&6.405&8.139$\times10^{-4}$&1.261$\times10^{-1}$\\
10&5&27.35&9.005$\times10^{-4}$&1.248$\times10^{0}$\\
5&1&1.155&4.641$\times10^{-4}$&9.652$\times10^{-4}$\\
5&2&1.485&5.899$\times10^{-4}$&3.018$\times10^{-3}$\\
5&3&2.512&7.134$\times10^{-4}$&1.589$\times10^{-2}$\\
5&4&6.441&8.186$\times10^{-4}$&1.269$\times10^{-1}$\\
5&5&27.40&9.018$\times10^{-4}$&1.250$\times10^{0}$\\
3&1&1.239&6.062$\times10^{-4}$&1.381$\times10^{-3}$\\
3&2&1.608&6.866$\times10^{-4}$&3.733$\times10^{-3}$\\
3&3&2.681&7.69$\times10^{-4}$&1.764$\times10^{-2}$\\
3&4&6.695&8.464$\times10^{-4}$&1.328$\times10^{-1}$\\
3&5&27.89&9.143$\times10^{-4}$&1.274$\times10^{0}$\\
\hline\hline
\end{tabular}
\label{table:mass}
\end{center}
\end{table}

\subsection{Thin Shell Approximation}

After the ejecta start expanding and interacting with the ambient medium, the hydrodynamical interaction of the two media leads to the formation of a couple of shock waves, forward and reverse shocks, when the pre-shocked pressure is sufficiently low at the interface between the two media as assumed in this study. 

As we have mentioned in Section \ref{sec:intro}, self-similar solutions describing the dynamical evolution of the shocked gas have been found in non-relativistic and ultra-relativistic regimes. 
However, we cannot expect any analytical solution for the trans-relativistic case, because the shock jump conditions at the two shock fronts cannot be simplified into convenient forms. 
In other words, the characteristic variables of the shocked gas cannot be expressed as simple power-law functions of time $t$ unlike the non-relativistic and ultra-relativistic cases. 
Despite the difficulty, the shocked gas can be regarded as a geometrically thin shell at early stages of the evolution as we will see below. 
Thus, we approximate  the width of the shell to be sufficiently small (referred to as ``thin shell approximation'') compared with the radius and solve the equation of motion of the shell. 
We further assume that the rest-mass energy of the shell dominates over the internal energy, $p/\rho\ll 1$. 
The validity of these approximations will be checked in Section \ref{sec:numerical}, where approximate solutions are compared with results of numerical simulations.

\subsection{Temporal Evolution of the Shell}

The dynamical evolution of the shell is determined by the following competing effects, the deceleration by loading mass, the supply of momentum through the shocks, and the deceleration due to the difference in the post-shock pressure, $p_\mathrm{fs}$ and $p_\mathrm{rs}$, at the forward and reverse shocks. 
We denote the mass, the Lorentz factor, and the velocity of the shell by $M_\mathrm{s}$, $\Gamma_\mathrm{s}$, $\beta_\mathrm{s}$, while the positions and the velocities of the forward and reverse shocks are $R_\mathrm{fs}$, $\beta_\mathrm{fs}$, $R_\mathrm{rs}$, and $\beta_\mathrm{rs}$. 
We model the temporal evolution of the momentum $S_{r}$ of the shell along the radial direction as follows,
\begin{equation}
\frac{dS_r}{d t}+
4\pi R_\mathrm{fs}^2F_\mathrm{fs}-4\pi R_\mathrm{rs}^2F_\mathrm{rs}=4\pi R_\mathrm{fs}^2p_\mathrm{rs}-4\pi R_\mathrm{rs}^2p_\mathrm{fs},
\label{eq:eom}
\end{equation}
The 1st and 2nd terms of the R.H.S. of Equation (\ref{eq:eom}) represent the force exerted by the post-shock pressure at the reverse and forward shock fronts. 
The quantities $-F_\mathrm{fs}$ and $F_\mathrm{rs}$ in the L.H.S. of Equation (\ref{eq:eom}) denote the momentum fluxes of gas flowing into the shocked region through the forward and the reverse shocks, which reflects the momentum conservation. 
Since the ambient medium is moving at a velocity much smaller than the ejecta, we neglect the contribution from the ambient medium to the momentum gain of the shell and thus set the corresponding flux to zero, $F_\mathrm{fs}=0$. 
On the other hand, the ejecta predominantly supply the shell with the momentum through the reverse shock. 
The corresponding momentum flux $F_\mathrm{rs}$ is described as follows,
\begin{equation}
F_\mathrm{rs}=\rho_\mathrm{ej,rs}\Gamma_\mathrm{ej,rs}^2\beta_\mathrm{ej,rs}(\beta_\mathrm{ej,rs}-\beta_\mathrm{rs}),
\end{equation}
where $\rho_\mathrm{ej,rs}$, $\Gamma_\mathrm{ej,rs}$, and $\beta_\mathrm{ej,rs}$ are the density, the Lorentz factor and the velocity of the pre-shocked ejecta at the reverse shock front.  

Next, the temporal evolution of the mass of the shell can be treated in a similar way to the momentum. 
The mass of the shocked gas continuously increases as the forward and reverse shocks sweep the ambient medium and the ejecta. 
The governing equation is expressed as follows,
\begin{equation}
\frac{dM_\mathrm{s}}{dt}+4\pi R_\mathrm{fs}^2G_\mathrm{fs}-4\pi R_\mathrm{rs}^2G_\mathrm{rs}=0.
\label{eq:diff_mass}
\end{equation}
The mass fluxes $-G_\mathrm{fs}$ and $G_\mathrm{rs}$ flowing into the shocked region through the forward and reverse shocks are expressed as follows,
\begin{equation}
G_\mathrm{fs}=\rho_\mathrm{a,fs}\Gamma_\mathrm{a,fs}(\beta_\mathrm{a,fs}-\beta_\mathrm{fs})=
-\rho_\mathrm{a,fs}\beta_\mathrm{fs},
\label{eq:Fm,fs}
\end{equation}
and
\begin{equation}
G_\mathrm{rs}=\rho_\mathrm{ej,rs}\Gamma_\mathrm{ej,rs}(\beta_\mathrm{ej,rs}-\beta_\mathrm{rs}).
\label{eq:Fm,rs}
\end{equation}
Here, $\rho_\mathrm{a,fs}$, $\Gamma_\mathrm{a,fs}$, and $\beta_\mathrm{a,fs}$ denote the density, the Lorentz factor, and the velocity of the pre-shocked ambient medium at the forward shock. 
Since the velocity of the ambient medium is negligibly small compared to the velocity of the shell, it is set to zero, $\beta_\mathrm{a,fs}=0$, and then the R.H.S of Equation (\ref{eq:Fm,fs}) is obtained. 

The momentum $S_r$ is expressed in terms of $M_\mathrm{s}$, $\Gamma_\mathrm{s}$, and $\beta_\mathrm{s}$ as follows,
\begin{equation}
S_r=M_\mathrm{s}\Gamma_\mathrm{s}\beta_\mathrm{s},
\end{equation}
which can be solved with respect to the Lorentz factor for a given set of the momentum and the mass,
\begin{equation}
\Gamma_\mathrm{s}=\sqrt{1+\frac{S_r^2}{M_\mathrm{s}^2}}.
\label{eq:Gamma_s}
\end{equation}

The radius of the shell evolves according to the following equation,
\begin{equation}
\frac{dR_\mathrm{s}}{dt}=\beta_\mathrm{s}.
\label{eq:diff_Rs}
\end{equation}
The temporal evolution of the forward and reverse shock radii, $R_\mathrm{fs}$ and $R_\mathrm{rs}$, are governed by similar equations,
\begin{equation}
\frac{dR_\mathrm{fs}}{dt}=\beta_\mathrm{fs},
\label{eq:diff_Rfs}
\end{equation}
and
\begin{equation}
\frac{dR_\mathrm{rs}}{dt}=\beta_\mathrm{rs}.
\label{eq:diff_Rrs}
\end{equation}
These equations can be integrated in a straightforward way, once the velocities, $\beta_\mathrm{s}$, $\beta_\mathrm{fs}$, and $\beta_\mathrm{rs}$, are obtained.

\subsection{Shock Jump Conditions}
The shock velocities, $\beta_\mathrm{rs}$ and $\beta_\mathrm{fs}$, and the hydrodynamical variables of the post-shock gas, $\rho_\mathrm{fs}$, $p_\mathrm{fs}$, $\rho_\mathrm{rs}$, and $p_\mathrm{rs}$, at the forward and reverse shocks appears in the governing equations introduced in the previous section.
To integrate the governing equations, these quantities should be calculated for a given set of the shock radii, $R_\mathrm{fs}$ and $R_\mathrm{rs}$ and the velocity of the shell $\beta_\mathrm{s}$, or the corresponding Lorentz factor $\Gamma_\mathrm{s}$. 
The shock jump conditions give these quantities. 
The derivation of the shock jump condition is described in detail elsewhere (see, Appendix of the previous paper, \cite{2014ApJ...796...30S}, some textbooks, or review papers, such as \cite{landau,MM2003}) and we will not repeat the derivation in this paper. 

The shock velocity $\beta_\mathrm{sh}$ is generally expressed as a function of the velocity $\beta_\mathrm{u}$ and the Lorentz factor $\Gamma_\mathrm{u}$ of the gas in the upstream and those, $\beta_\mathrm{d}$ and $\Gamma_\mathrm{d}$, in the downstream as follows,
\begin{eqnarray}
&&\beta_\mathrm{sh}(\beta_\mathrm{u},\beta_\mathrm{d})
\nonumber\\
&&\hspace{1em}=
\frac{
\gamma\Gamma_\mathrm{u}\Gamma_\mathrm{d}^2(\beta_\mathrm{u}-\beta_\mathrm{d})\beta_\mathrm{d}
-(\gamma-1)(\Gamma_\mathrm{u}-\Gamma_\mathrm{d})
}
{
\gamma\Gamma_\mathrm{u}\Gamma_\mathrm{d}^2(\beta_\mathrm{u}-\beta_\mathrm{d})
-(\gamma-1)(\Gamma_\mathrm{u}\beta_\mathrm{u}-\Gamma_\mathrm{d}\beta_\mathrm{d})
},
\end{eqnarray}
when the pressure in the upstream is negligible. 
Once the shock velocity $\beta_\mathrm{sh}$ is obtained and the pre-shock density $\rho_\mathrm{u}$ is known, the post-shock density $\rho_\mathrm{d}$ and pressure $p_\mathrm{d}$ are found from the following relations,
\begin{equation}
\rho_\mathrm{d}=\rho_\mathrm{u}
\frac{
\Gamma_\mathrm{u}(\beta_\mathrm{u}-\beta_\mathrm{sh})
}
{
\Gamma_\mathrm{d}(\beta_\mathrm{d}-\beta_\mathrm{sh})
},
\end{equation} 
and
\begin{equation}
p_\mathrm{d}=
\frac{\rho_\mathrm{u}
\Gamma_\mathrm{u}^2(\beta_\mathrm{u}-\beta_\mathrm{d})(\beta_\mathrm{u}-\beta_\mathrm{sh})
}
{
1-\beta_\mathrm{d}\beta_\mathrm{sh}
},
\end{equation} 
where $\Gamma_\mathrm{sh}=(1-\beta_\mathrm{sh}^2)^{-1/2}$ is the shock Lorentz factor. 

\subsubsection{Forward Shock}
For the forward shock, the pre-shock velocity is equal to zero and the pre-shock density $\rho_\mathrm{a,fs}$ is obtained by substituting the forward shock radius $R_\mathrm{fs}$ into the density profile of the ambient medium, Equation (\ref{eq:csm}),
\begin{equation}
\rho_\mathrm{a,fs}=AR^{-2}_\mathrm{fs}.
\end{equation}
Thus, regarding the shell velocity $\beta_\mathrm{s}$ as the post-shock velocity, the forward shock velocity $\beta_\mathrm{fs}$ is obtained as follows,
\begin{equation}
\beta_\mathrm{fs}=\beta_\mathrm{sh}(0,\beta_\mathrm{s}).
\label{eq:beta_fs}
\end{equation}
Using the shock velocity, the forward shock Lorentz factor $\Gamma_\mathrm{fs}$, the post-shock density $\rho_\mathrm{fs}$, and the post-shock pressure $p_\mathrm{fs}$ can be calculated as follows,
\begin{equation}
\Gamma_\mathrm{fs}=\frac{1}{\sqrt{1-\beta_\mathrm{fs}^2}},
\end{equation}
\begin{equation}
\rho_\mathrm{fs}=\rho_\mathrm{a,fs}
\frac{
\beta_\mathrm{fs}
}
{
\Gamma_\mathrm{s}(\beta_\mathrm{fs}-\beta_\mathrm{s})
},
\label{eq:rho_fs,d}
\end{equation} 
and
\begin{equation}
p_\mathrm{fs}=
\rho_\mathrm{a,fs}
\frac{
\beta_\mathrm{s}\beta_\mathrm{fs}
}
{
1-\beta_\mathrm{s}\beta_\mathrm{fs}
}.
\label{eq:pre_fs,d}
\end{equation} 
\subsubsection{Reverse shock}
The reverse shock propagates in the freely expanding ejecta. 
Thus, the pre-shock values, $\beta_\mathrm{ej,rs}$ and $\rho_\mathrm{ej,rs}$, of the velocity and the density are given by
\begin{equation}
\beta_\mathrm{ej,rs}=\frac{R_\mathrm{rs}}{t},
\end{equation}
and
\begin{equation}
\rho_\mathrm{ej,rs}=\rho_\mathrm{ej}(t,R_\mathrm{rs}).
\end{equation}
The reverse shock velocity is determined in a similar way to the forward shock,
\begin{equation}
\beta_\mathrm{rs}=\beta_\mathrm{sh}(\beta_\mathrm{ej,rs},\beta_\mathrm{s}).
\label{eq:beta_rs}
\end{equation}
Once the reverse shock velocity is obtained, the reverse shock Lorentz factor $\Gamma_\mathrm{rs}$, the post-shock density $\rho_\mathrm{rs}$, and the post-shock pressure $p_\mathrm{rs}$ are evaluated as follows,
\begin{equation}
\Gamma_\mathrm{rs}=\frac{1}{\sqrt{1-\beta_\mathrm{rs}^2}},
\end{equation}
\begin{equation}
\rho_\mathrm{rs}=\rho_\mathrm{ej,rs}
\frac{
\Gamma_\mathrm{ej,rs}(\beta_\mathrm{ej,rs}-\beta_\mathrm{rs})
}
{
\Gamma_\mathrm{u}(\beta_\mathrm{s}-\beta_\mathrm{rs})
},
\label{eq:rho_rs,d}
\end{equation} 
and
\begin{equation}
p_\mathrm{rs}=
\frac{\rho_\mathrm{ej,rs}\Gamma_\mathrm{ej,rs}^2(\beta_\mathrm{ej,rs}-\beta_\mathrm{rs})
(\beta_\mathrm{ej,rs}-\beta_\mathrm{s})}
{1-\beta_\mathrm{s}\beta_\mathrm{rs}}.
\label{eq:pre_rs,d}
\end{equation} 

\subsection{Non-relativistic and Ultra-relativistic Limits}
When the velocity of the shell is much smaller than the speed of light (``non-relativistic regime'') or the Lorentz factor of the shell is much larger than unity (``ultra-relativistic regime''), self-similar solutions describing the flow are known \citep{1982ApJ...259..302C,2006ApJ...645..431N}. 
In the following, we consider the relation between our model and the non-relativistic and ultra-relativistic cases by reproducing the temporal behavior of the shell in these two limits. 

\subsubsection{Non-relativistic Regime}
The density profile in the non-relativistic limit is obtained by setting $\Gamma=1$ in Equation (\ref{eq:ejecta_profile}), 
\begin{equation}
\rho_\mathrm{ej,NR}\propto t^{-3}\beta^{-n}.
\end{equation}
The dependence of the shell velocity $\beta_\mathrm{s}$ on time $t$ is same as the shell radius divided by time $t$, $\beta_\mathrm{s} \propto R_\mathrm{s}/t$. 
Since the flow is self-similar, the forward and reverse shock velocities are proportional to that of the shell. 
Therefore, the pressure of the gas in the downstream of the reverse shock, Equation (\ref{eq:pre_rs,d}), should satisfy
\begin{equation}
p_\mathrm{rs,NR}\propto t^{-3}\beta_\mathrm{s}^{-n+2},
\end{equation}
under the limit of $\beta_\mathrm{s},\beta_\mathrm{rs},\beta_\mathrm{ej,rs}\ll 1$. 
On the other hand, Equation (\ref{eq:pre_fs,d}) gives the following relation for the gas pressure in the downstream of the forward shock,
\begin{equation}
p_\mathrm{fs,NR}\propto t^{-k}\beta_\mathrm{s}^{-k+2}.
\end{equation}

The time dependence of physical variables of the flow can be obtained by imposing the condition that the dependence of the post-shock pressure at the forward and reverse shock fronts on the time $t$ should be identical, $p_\mathrm{fs}/p_\mathrm{rs}\propto t^0$. 
The condition $p_\mathrm{fs,NR}/p_\mathrm{rs,NR}\propto t^0$ gives the time dependence of the velocity $\beta_\mathrm{s}$,
\begin{equation}
\beta_\mathrm{s,NR}\propto t^{(k-3)/(n-k)}.
\end{equation}
Thus, the radius of the shell evolves as follows,
\begin{equation}
R_\mathrm{s,NR}\propto \beta_\mathrm{s,NR}t\propto t^{(n-3)/(n-k)},
\end{equation}
which agrees with the dependence derived by \cite{1982ApJ...259..302C}. 

\subsubsection{Ultra-relativistic Regime}
In the ultra-relativistic regime, Equations (\ref{eq:pre_fs,d}) and (\ref{eq:pre_rs,d}) lead to,
\begin{equation}
p_\mathrm{fs,UR}\propto t^{-k}\Gamma_\mathrm{s}^2,
\end{equation}
and
\begin{equation}
p_\mathrm{rs,UR}\propto t^{-3}\Gamma_\mathrm{s}^{-n}.
\end{equation}
Therefore, one obtains the following time dependence of the Lorentz factor,
\begin{equation}
\Gamma_\mathrm{s,UR}\propto t^{-(3-k)/(n+2)},
\label{eq:gamma_ur}
\end{equation}
by using the condition $p_\mathrm{fs,UR}/p_\mathrm{rs,UR}\propto t^0$. 
This is consistent with the time dependence of the Lorentz factor in the self-similar solution found by \cite{2006ApJ...645..431N}.

\subsection{Initial Velocity of the Shell}
One has to evaluate the initial velocity of the shell to integrate Equations (\ref{eq:eom}), (\ref{eq:diff_mass}), (\ref{eq:diff_Rs}), (\ref{eq:diff_Rfs}), and (\ref{eq:diff_Rrs}). 
The outermost layer of the freely expanding ejecta is initially adjacent to the ambient medium at $r=\beta_\mathrm{max}t_0$. 
At the very beginning of the hydrodynamical interaction, the forward and reverse shock radii are identical with the interface, $r=\beta_\mathrm{max}t_0$, at which the following pressure balance is achieved in the shell,
\begin{equation}
p_\mathrm{fs}=p_\mathrm{rs}.
\end{equation}
By evaluating the pre-shocked density and the velocity of the ejecta and the ambient medium at $r=\beta_\mathrm{max}t_0$, the pressure $p_\mathrm{fs}$ and $p_\mathrm{rs}$ can be calculated for a given value of the shell velocity $\beta_\mathrm{s}$. 
Thus, the initial value of the shell velocity can be determined as the velocity satisfying the above pressure condition for a given set of the time $t_0$, the maximum velocity $\beta_\mathrm{max}$, the parameter $A_\star$, the exponent $n$, and the kinetic energy $E_\mathrm{rel}$. 
In fact, this procedure is identical with solving a shock tube problem of special relativistic hydrodynamics. 

Once the initial velocity is obtained, Equations (\ref{eq:eom}), (\ref{eq:diff_mass}), (\ref{eq:diff_Rs}), (\ref{eq:diff_Rfs}), and (\ref{eq:diff_Rrs}) can be integrated. 
For a given set of the variables $M_\mathrm{s}(t)$, $\beta_\mathrm{s}(t)$, and $R_\mathrm{s}(t)$ at $t$, their values at the next time step $t+\Delta t$ are explicitly obtained from the equations discretized in time. 
Then, Equation (\ref{eq:Gamma_s}) is used to find the Lorentz factor of the shell.

Results of the integration for several sets of the free parameters are shown in the next section and compared with one-dimensional hydrodynamic simulations. 

\section{NUMERICAL SIMULATIONS}\label{sec:numerical}
We perform simulations of the hydrodynamical interaction for several sets of the free parameters by numerically solving the equations of special relativistic hydrodynamics, Equations (\ref{eq:continuity}) - (\ref{eq:enthalpy}). 
Our method to integrate the equations is described in the previous paper \citep{2014ApJ...796...30S}. 

The radial coordinate of the numerical domain ranges from $r=10^8$ cm to $r=4\times 10^{13}$ cm. 
An adaptive mesh refinement (AMR) technique is implemented in the code to efficiently capture discontinuities. 
The whole numerical domain is covered by 1024 cells (``base grid'') with the AMR refinement level of $l=0$. 
The size of a cell is halved when the cell needs a finer resolution. 
The maximum refinement level is set to $l_\mathrm{max}=12$, which gives a resolution finer by a factor of $2^{12}=4096$ than the base grid. 

\begin{figure}[tbp]
\begin{center}
\includegraphics[scale=0.45]{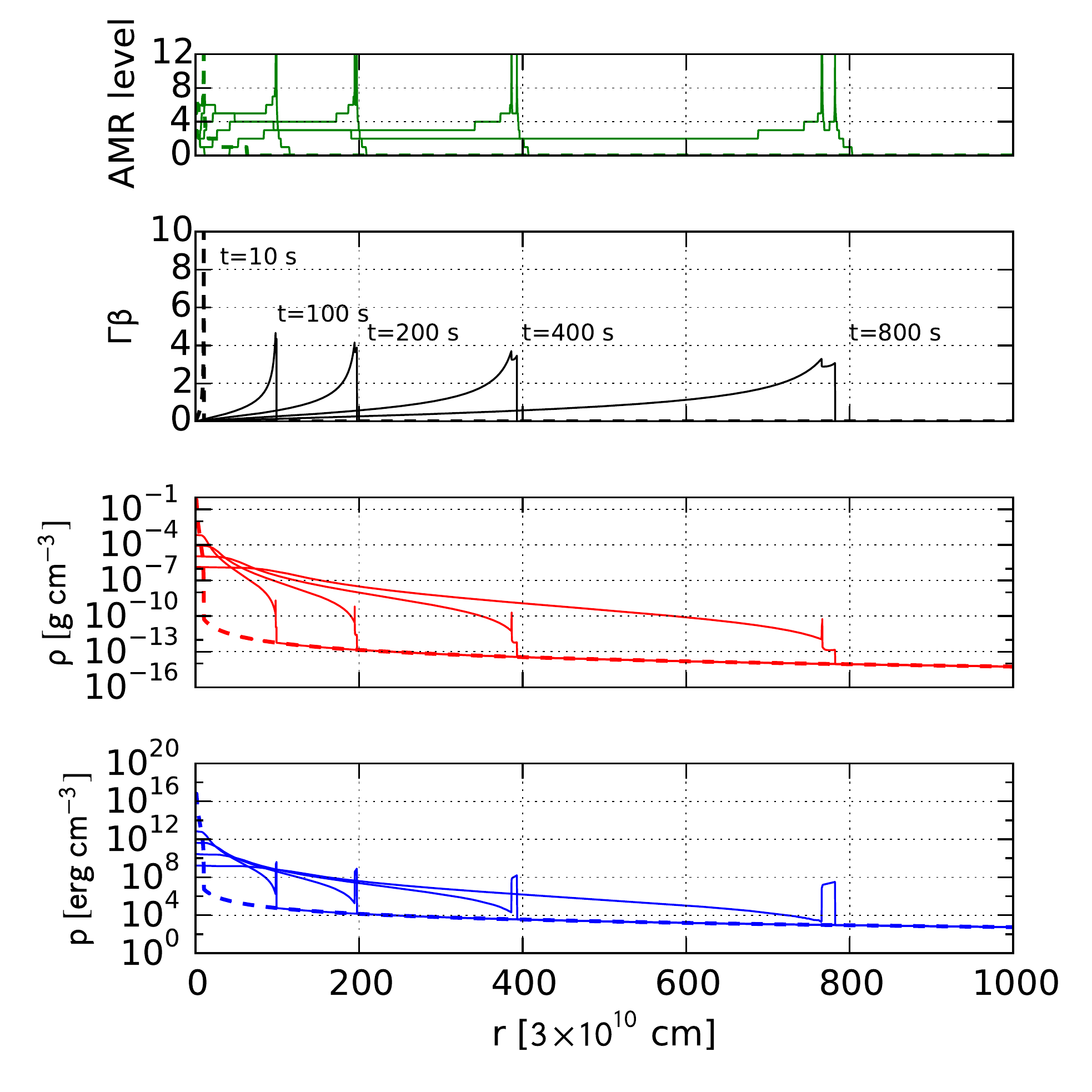}
\caption{Radial profiles of the AMR refinement level, the Lorentz factor, the density, and the pressure at $t=100$, $200$, $400$, and $800$ s. 
The parameters of the ejecta and the ambient medium are set to be $t_0=10$ s, $n=4$, $E_\mathrm{rel,51}=1.0$, $\Gamma_\mathrm{max}=10$, and $A_{\star}=10$. }
\label{fig:evolution}
\end{center}
\end{figure}

\begin{figure*}[tbp]
\begin{center}
\includegraphics[scale=0.35]{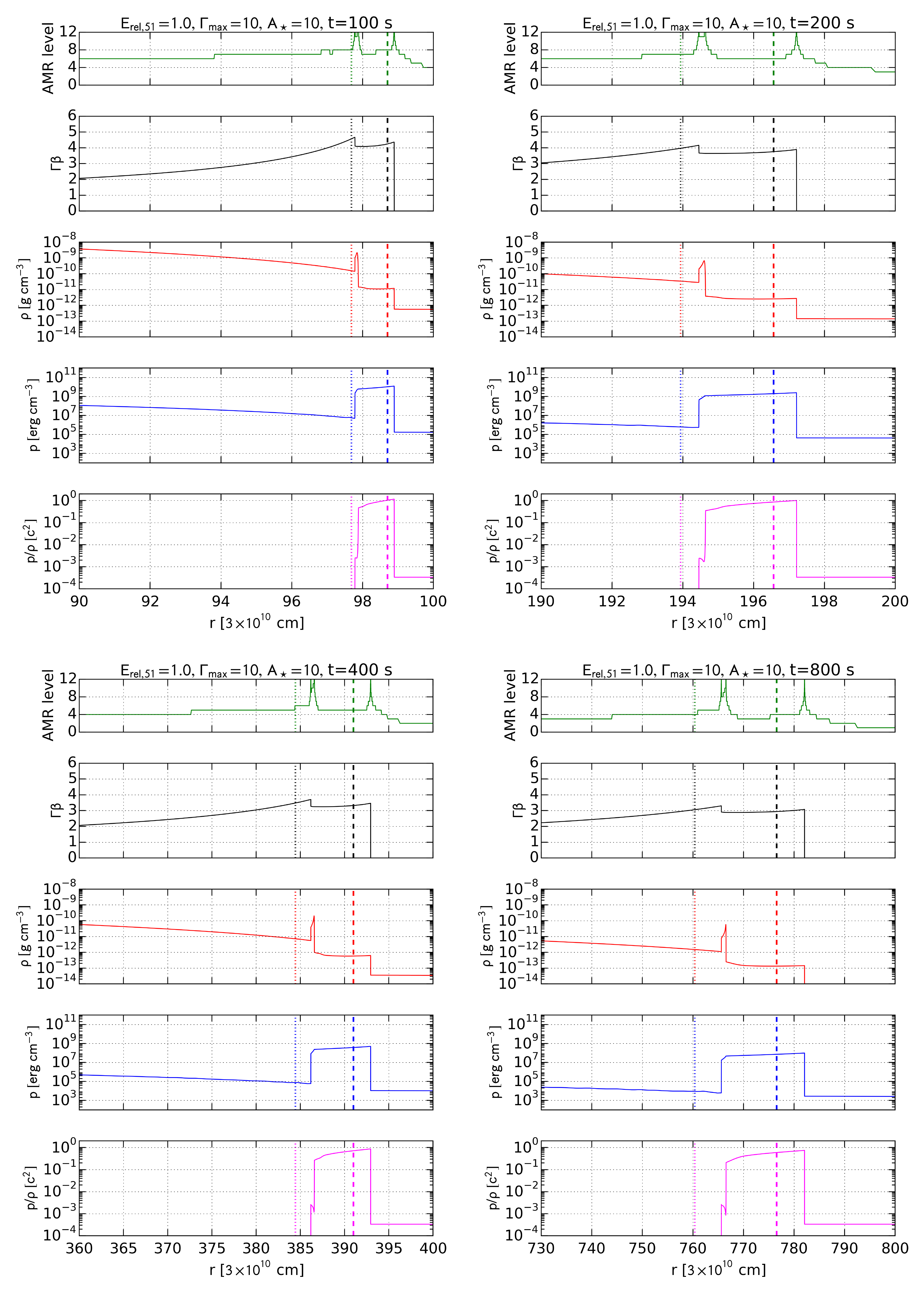}
\caption{Enlarged views of the radial profiles shown in Figure \ref{fig:evolution}. 
The forward shock in the ambient medium, the reverse shock in the ejecta, and the contact discontinuity separating the shocked ambient medium and ejecta are clearly resolved. 
The bottom panel in each snapshot shows the profile of $p/\rho$. 
The dashed and dotted vertical lines in each panel correspond to the positions of the forward and reverse shock fronts predicted by the semi-analytical model. }
\label{fig:profile}
\end{center}
\end{figure*}

\subsection{Initial and Boundary Conditions}
The following initial conditions are imposed on the density, velocity, and pressure profiles. 
The ejecta and the ambient medium are initially separated at $r=\beta_\mathrm{max}t_0$. 
The density and the velocity profiles at $t=t_0$ are given by
\begin{equation}
\rho(t_0,r)=\left\{\begin{array}{ccl}
\rho_\mathrm{ej}(t_0,r)&\mathrm{for}&r\leq\beta_\mathrm{max}t_0,\\
\rho_\mathrm{a}(r)&\mathrm{for}&r>\beta_\mathrm{max}t_0,
\end{array}\right.
\end{equation}
and
\begin{equation}
\beta(t_0,r)=\left\{\begin{array}{ccl}
\frac{r}{t_0}&\mathrm{for}&r\leq\beta_\mathrm{max}t_0,\\
0&\mathrm{for}&r>\beta_\mathrm{max}t_0.
\end{array}\right.
\end{equation}
We assume the following pressure profile,
\begin{equation}
p(t_0,r)=10^{-3}\rho(t_0,r),
\end{equation}
so that the pressure does not affect the dynamical evolution of the ejecta.
At the inner and outer boundaries, free boundary conditions are imposed. 

\subsection{Results}

We carry out simulations with several sets of the free parameters. 
The initial time $t_0$ and the exponent $n$ are fixed to be $t_0=10.0$ s and $n=4$ in all calculations. 
In order to see the dependence of the evolution of the shell on the maximum Lorentz factor, we calculate models with different maximum Lorentz factors $\Gamma_\mathrm{max}=10$, $5$, and $3$, while the kinetic energy and the ambient density are fixed, $E_\mathrm{rel,51}=1.0$ and $A_\star=10$. 
In addition, we calculate models with a larger ambient density $A_\star=1000$ and different kinetic energies $E_\mathrm{rel,51}=1.0$, $0.1$, and $0.01$. 
In the latter models, the shell effectively decelerates by sweeping the dense ambient medium and a transition from relativistic to non-relativistic speeds is realized. 

Figures \ref{fig:evolution} and \ref{fig:profile} show results of the simulation with $E_\mathrm{rel,51}=1.0$, $\Gamma_\mathrm{max}=10$, and  $A_\star=10$. 
In Figure \ref{fig:evolution}, the radial profiles of the Lorentz factor, the density, and the pressure at $t=10,\ 100,$ $200$, $400$, and $800$ s are shown. 
The AMR refinement level is also plotted as a function of the radial coordinate in the top panel of Figure \ref{fig:evolution}. 
Immediately after the simulation starts, a shell composed of the shocked ejecta and the shocked ambient medium forms around the interface between the two media. 
The width of the shell is much smaller than its radius, ensuring the validity of the thin shell approximation adopted in the previous section. 
The profiles of these variables around the shock fronts are presented in Figure \ref{fig:profile}. 
In addition to the quantities shown in Figure \ref{fig:evolution}, the radial profiles of $p/\rho$, which is proportional to the specific internal energy, are also plotted. 
The two shock fronts and the contact discontinuity separating the shocked ejecta and the shocked ambient medium are clearly resolved thanks to the AMR technique.

\begin{figure*}[tbp]
\begin{center}
\includegraphics[scale=0.38]{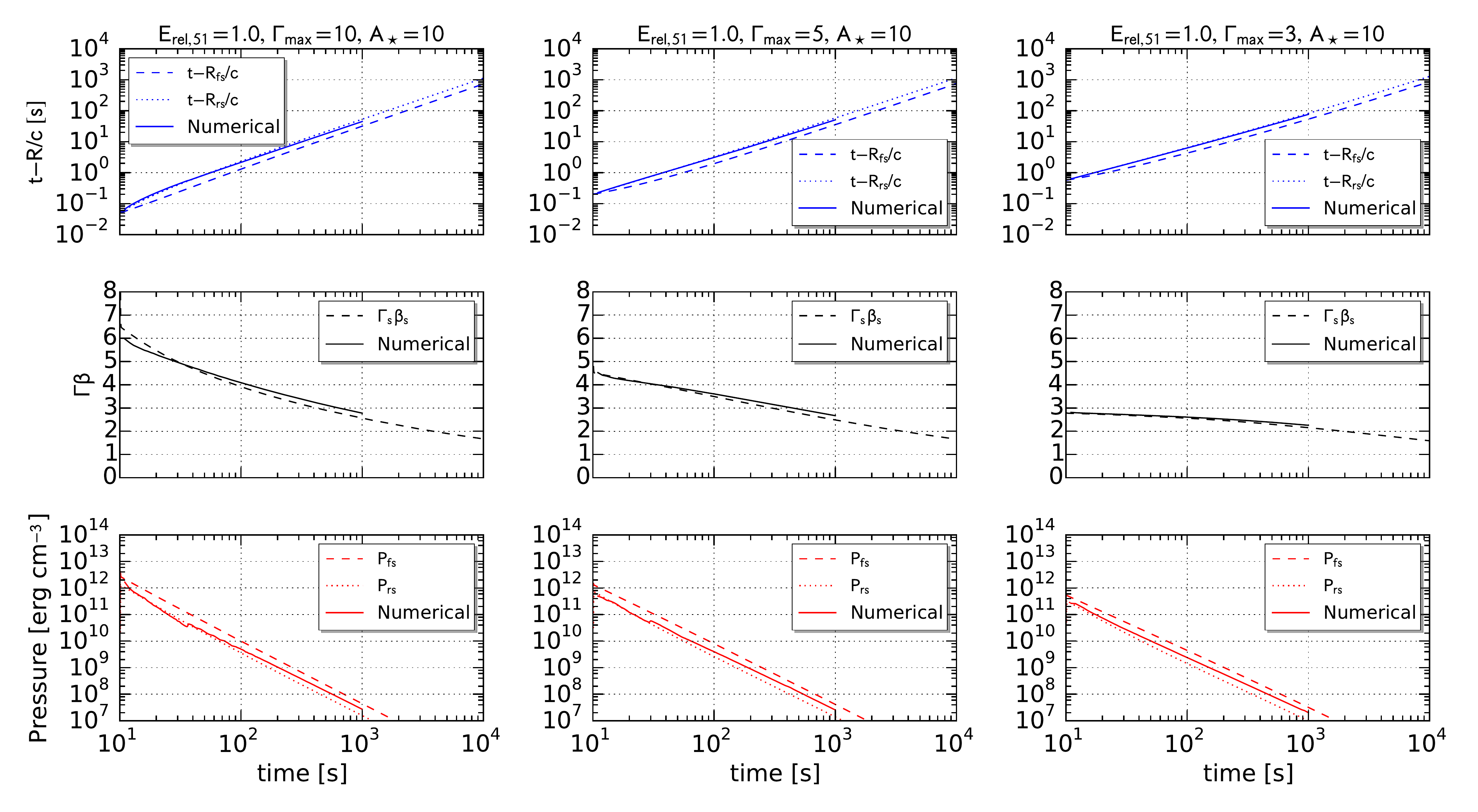}
\caption{Temporal evolutions of the shock radii, the Lorentz factor, and the pressure. 
Models with different maximum Lorentz factors $\Gamma_\mathrm{max}=10$ (left), $5$ (center), and $3$ (right) and fixed values of the kinetic energy $E_\mathrm{rel,51}=1.0$, the exponent $n=4$, the initial time $t_0=10$ s, and the ambient density $A_\star=10$ are shown. 
In each model, the evolutions of the shock radii or the shell radius (top), the 4-velocity of the shell (middle), and the pressure (bottom) are presented. 
In all panels, the solid lines show these quantities obtained from numerical simulations, while the dashed and dotted lines show those of the semi-analytical model. 
In the upper panels, the dashed and dotted lines correspond to the forward and reverse shocks. 
In the middle panels, the dashed line correspond to the 4-velocity of the shell. 
In the bottom panels, the dashed and dotted lines show the post-shock pressures at the forward and reverse shock fronts.}
\label{fig:scaling1}
\end{center}
\end{figure*}

\begin{figure*}[tbp]
\begin{center}
\includegraphics[scale=0.38]{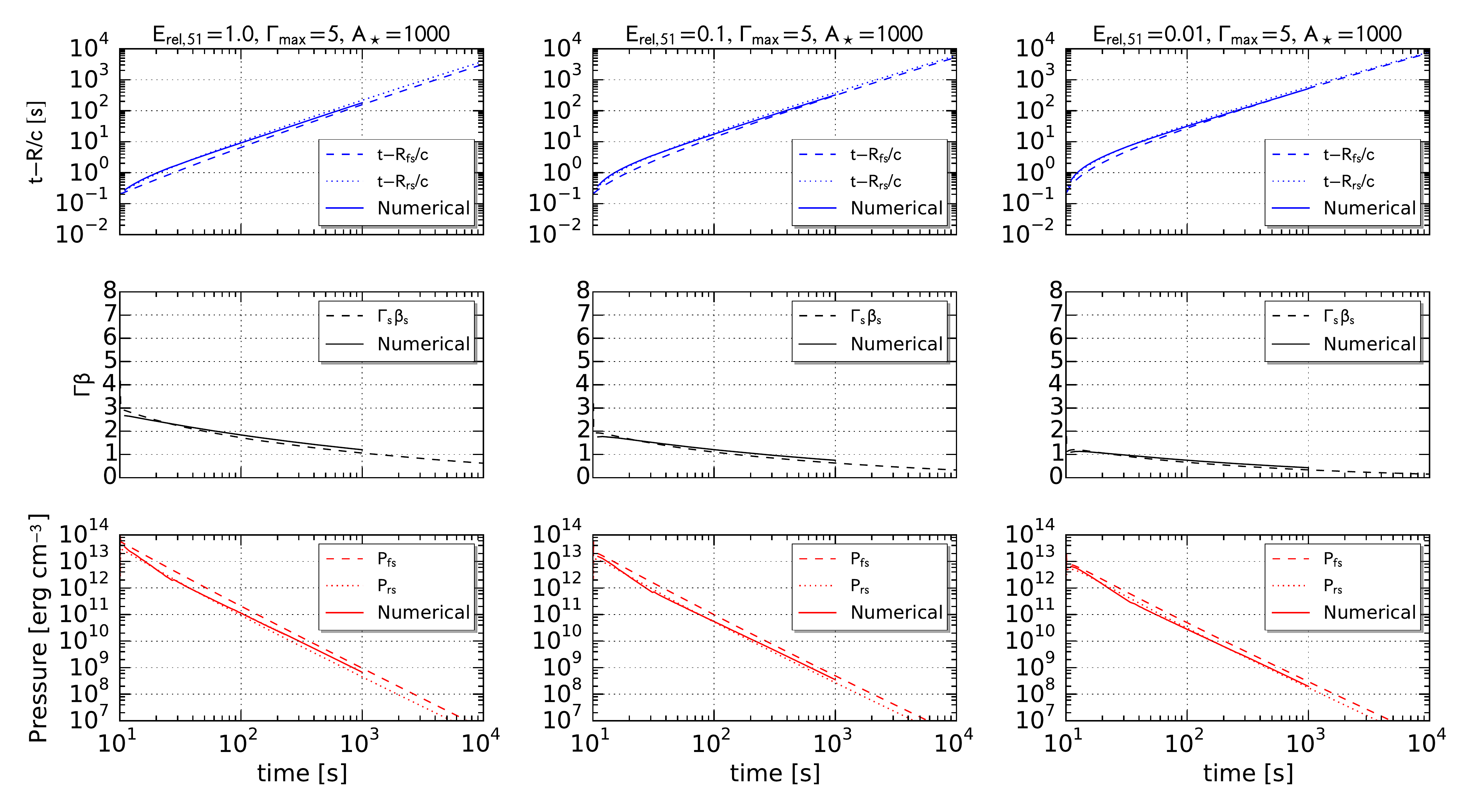}
\caption{Same as Figure \ref{fig:scaling1}, but for models with different kinetic energies $E_\mathrm{rel,51}=1.0$ (left), $0.1$ (center), and $0.01$ (right) and a relatively high ambient density $A_\star=1000$. 
}
\label{fig:scaling2}
\end{center}
\end{figure*}

The positions of the forward and reverse shock fronts, the 4-velocity of the shell, and the post-shock pressures at the forward and reverse shock fronts at time $t$ are obtained by the semi-analytical model. 
In each panel of Figure \ref{fig:profile}, the positions of the forward and reverse shock fronts are presented as dashed and dotted vertical lines. 
The semi-analytical model predicts slightly smaller shock radii. 
However, the differences are not significant. 
In Figures \ref{fig:scaling1} and  \ref{fig:scaling2}, the temporal evolutions of these quantities are plotted. 
For the shock radii $R_\mathrm{fs}$ and $R_\mathrm{rs}$, we plot the differences, $t-R_\mathrm{fs}/c$ and $t-R_\mathrm{rs}/c$, between $r=ct$ and the radii divided by the speed of light. 
The dashed line in the 4-velocity evolution shows the 4-velocity of the shell in the semi-analytical calculation and the dashed and dotted lines in the pressure evolution show the forward and reverse shock pressures. 
The solid line in each panel shows a good agreement with dashed and dotted lines, suggesting that the semi-analytical method developed in the previous section well describes the dynamical evolution of the shell resulting from the ejecta-ambient medium interaction. 

In Figure \ref{fig:scaling1}, models with different maximum Lorentz factors $\Gamma_\mathrm{max}=10$, $5$, and $3$ are compared. 
Although the initial 4-velocity of the shell can differ from each other, these models exhibit similar evolutions after the high Lorentz factor component has been swept and the 4-velocity reaches a few, $\Gamma\beta\simeq 2$. 
Thus, the maximum Lorentz factor of the shell does not influence the dynamical evolution at later epochs. 
In Figure \ref{fig:scaling2}, we show models in which the shell efficiently decelerates due to the small kinetic energy and the high ambient density. 
Even when the 4-velocity becomes smaller than unity, the semi-analytical model works well.

\subsection{Comparison of Semi-analytical Model with Numerical Results and Self-similar solutions}

The temporal evolution of the variables can also be compared with that of the self-similar solution in the ultra-relativistic regime. 
Setting the exponents in Equation (\ref{eq:gamma_ur}) to $k=2$ and $n=4$, one finds that the Lorentz factor $\Gamma_\mathrm{s}$ and the post-shock pressure of the shocks evolve as,
\begin{equation}
\Gamma_\mathrm{s,UR}\propto t^{-0.167},\ \ \ 
p_\mathrm{fs,UR}\propto t^{-2.33},
\end{equation}
in the ultra-relativistic regime. 

We fit power-law functions of time $t$ to the Lorentz factor and the forward shock pressure of the semi-analytical model with $\Gamma_\mathrm{max}=10$ and find that they evolve as
\begin{equation}
\Gamma_\mathrm{s}\propto t^{-0.157},\ \ \ 
p_\mathrm{fs}\propto t^{-2.33}.
\end{equation}
The time dependence of the Lorentz factor deviates from the self-similar solution because the ejecta velocities are mildly relativistic.  
The deviation is larger for models with smaller maximum Lorentz factors. 
In fact, the temporal evolution of the Lorentz factor realized in the models with $\Gamma_\mathrm{max}=5$ and $3$ shown in Figure \ref{fig:scaling1} cannot be well fitted by a simple power-law function. 
This clearly demonstrates that the self-similar approach fails for mildly relativistic ejecta.

\section{EMISSION FROM THE SHOCKED GAS}\label{sec:emission}
In the previous section, it has been confirmed that the semi-analytical model well agrees with numerical simulations. 
Then, in this section, we use the semi-analytical model to investigate the expected emission from the shocked gas. 
We assume that the ejecta are produced by an explosion of a star having lost its hydrogen and helium layers and the stellar atmosphere is mainly composed of oxygen. 
Thus, we set the mass number and the atomic number of ions to $A_\mathrm{i}=16$ and $Z_\mathrm{i}=8$. 

\subsection{Temporal Evolution of the Shell}

The kinetic energy of the ejecta dissipated via the reverse shock could be a plausible source of high-energy emission from the shell. 
The rate of the dissipation is governed by the dynamical evolution of the shell. 
From the semi-analytical model, one can evaluate the internal energy of the shell at time $t$ in the following way. 

The internal energy $E_\mathrm{s}(t)$ of the shell evolves according to the following equation,
\begin{eqnarray}
&&\frac{dE_\mathrm{s}}{dt}+4\pi R_\mathrm{fs}^2H_\mathrm{fs}-4\pi R_\mathrm{rs}^2H_\mathrm{rs}
\nonumber\\
&&\hspace{6em}=-\frac{E_\mathrm{s}}{3V_\mathrm{s}}\frac{dV_\mathrm{s}}{dt}-\left(\frac{dE}{dt}\right)_\mathrm{rad},
\label{eq:dEdt}
\end{eqnarray}
where the energy fluxes $H_\mathrm{fs}$ and $H_\mathrm{rs}$ through the froward and reverse shock fronts are given by,
\begin{equation}
H_\mathrm{fs}=4p_\mathrm{fs}\Gamma_\mathrm{s}^2(\beta_\mathrm{s}-\beta_\mathrm{fs}),
\end{equation}
and
\begin{equation}
H_\mathrm{rs}=4p_\mathrm{rs}\Gamma_\mathrm{s}^2(\beta_\mathrm{s}-\beta_\mathrm{rs}). 
\end{equation}
Here, we have introduced the volume of the shell, 
\begin{equation}
V_\mathrm{s}=\frac{4\pi}{3}(R_\mathrm{fs}^3-R_\mathrm{rs}^3).
\end{equation}
The 2nd and 3rd terms in the L.H.S. of Equation (\ref{eq:dEdt}) describe the rate of the change in the internal energy of the shell per unit time due to the shock passage, while the 1st and 2nd terms in the R.H.S. describe the energy loss per unit time due to adiabatic expansion and radiative diffusion. 
We evaluate the adiabatic loss term, which reflects the work done by the shell, as follows,
\begin{equation}
\frac{E_\mathrm{s}}{3V_\mathrm{s}}\frac{dV_\mathrm{s}}{dt}=\frac{4\pi (R_\mathrm{fs}^2\beta_\mathrm{fs}-R_\mathrm{rs}^2\beta_\mathrm{rs})}{V_\mathrm{s}}E_\mathrm{s}. 
\end{equation}

As discussed in Appendix \ref{sec:diffusion}, the expression of the radiative diffusion term, Equation (\ref{eq:dEdt_rad}), can be obtained by considering diffusion of photons in a geometrically thin shell moving at a relativistic speed,
\begin{equation}
\left(\frac{dE}{dt}\right)_\mathrm{rad}=4\pi R_\mathrm{s}^2u_\mathrm{ph}\frac{1-\beta_\mathrm{s}^2}{(3+\beta_\mathrm{s}^2)\tau+2\beta_\mathrm{s}}.
\label{eq:dEdt2}
\end{equation}
We assume that the internal energy of the shell is dominated by radiation, i.e., the radiation energy density $u_\mathrm{ph}$ in Equation (\ref{eq:dEdt_rad}) is identical with the internal energy density $u_\mathrm{int}$ of the gas in the shell, $u_\mathrm{ph}=u_\mathrm{int}$. 
We calculate the value by dividing the internal energy $E_\mathrm{s}$ of the shell by the volume,
\begin{equation}
u_\mathrm{int}=\frac{E_\mathrm{s}}{V_\mathrm{s}}.
\label{eq:u_int}
\end{equation}
We will check the consistency of this assumption below. 
Furthermore, under the thin shell approximation, the total optical depth $\tau$ of the shell for electron scattering is given by,
\begin{equation}
\tau=\frac{\kappa_\mathrm{es}M_\mathrm{s}}{4\pi R_\mathrm{s}^2},
\end{equation}
where $\kappa_\mathrm{es}=0.2$ cm$^2$ g$^{-1}$ is the electron scattering opacity for hydrogen-free gas \cite[e.g.,][]{1979rpa..book.....R}. 
These equations can easily be integrated along with Equations (\ref{eq:eom}) and (\ref{eq:diff_mass}). 
Since the diffusion approximation used to derive the expression of $(dE/dt)_\mathrm{rad}$ is no longer valid after the shell becomes transparent to photons, $\tau<1$, we stop the integration when the optical depth of the shell becomes unity at $t=t_\mathrm{tr}$. 

\begin{figure}[tbp]
\begin{center}
\includegraphics[scale=0.55]{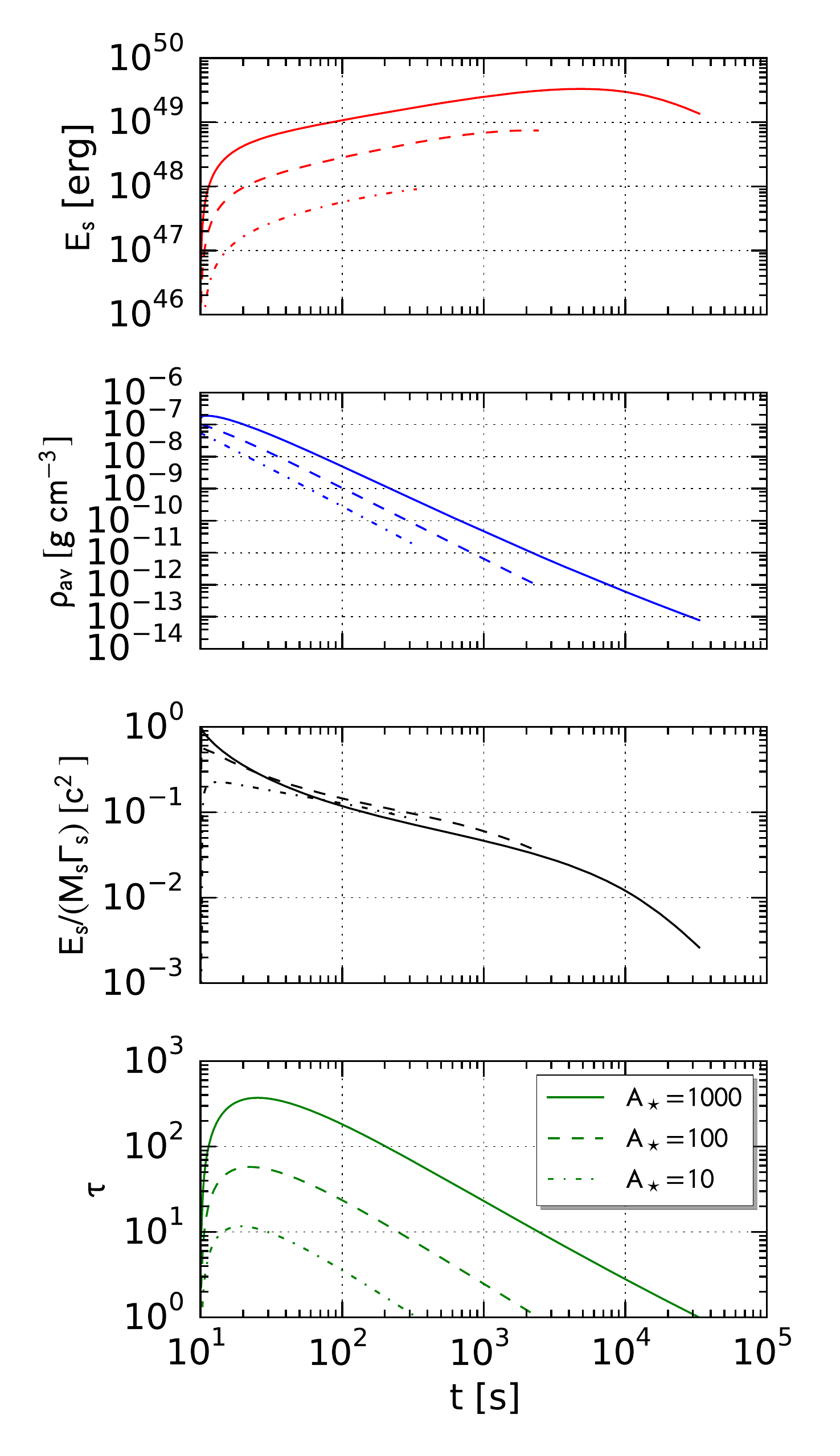}
\caption{Temporal evolutions of the internal energy, the average density, the average specific internal energy, and the total optical depth of the shell calculated by the semi-analytical model from top to bottom. 
Models with $A_{\star}=10$, $100$, and $1000$ are shown. 
The kinetic energy $E_\mathrm{rel}$, the time $t_0$, the maximum Lorentz factor $\Gamma_\mathrm{max}$, and the exponent $n$ are fixed to be $E_\mathrm{rel,51}=1.0$, $t_0=10$ s, $\Gamma_\mathrm{max}=5$, and $n=4$. }
\label{fig:emission}
\end{center}
\end{figure}

The temporal evolution of the internal energy $E_\mathrm{s}$, the average density $\rho_\mathrm{av}$, the average specific internal energy, and the optical depth $\tau$ of the shell for models with several values of the parameter $A_\star$ are shown in Figure \ref{fig:emission}. 
The average density of the shell is evaluated by dividing the mass $M_\mathrm{s}$ by the volume $V_\mathrm{s}$ and the Lorentz factor $\Gamma_\mathrm{s}$ of the shell, $\rho_\mathrm{av}=M_\mathrm{s}/(\Gamma_\mathrm{s}V_\mathrm{s})$. 
The average specific internal energy is calculated by dividing the internal energy $E_\mathrm{s}$ by the mass $M_\mathrm{s}$ and the Lorentz factor $\Gamma_\mathrm{s}$, $E_\mathrm{s}/(M_\mathrm{s}\Gamma_\mathrm{s})$. 
As shown the radial profiles of $p/\rho$ in Figure \ref{fig:profile}, the difference in the relative velocities of the forward and reverse shocks results in quite different values of the ratio $p/\rho$, which is proportional to the specific internal energy, of the shocked ambient gas and the shocked ejecta. 
Thus, it should be noted that the specific internal energy $E_\mathrm{s}/(M_\mathrm{s}\Gamma_\mathrm{s})$ only gives an averaged value over the entire shocked region. 
In the models, the time $t_0$, the kinetic energy $E_\mathrm{rel}$, the maximum Lorentz factor $\Gamma_\mathrm{max}$, and the exponent $n$ are fixed to $t_0=10$ s, $E_\mathrm{rel,51}=1$, $\Gamma_\mathrm{max}=5$, and $n=4$, while the parameter $A_\star$ is set to $A_{\star}=10$, $100$, and $1000$. 
For larger values of the parameter $A_\star$, the kinetic energy dissipation is more efficient and the time scale $t_\mathrm{tr}$ becomes longer. 
In early stages of its temporal evolution, the internal energy of the shell continues to increase as it is supplied by the reverse shock. 
For models with larger $A_\star$, the internal energy starts decreasing afterward, because the contribution of the radiative loss eventually becomes significant. 
For models with smaller $A_\star$, the shell becomes transparent to photons before the internal energy starts decreasing. 
The average density of the shell continues to decline almost throughout its evolution due to the expansion of the shell. 
The optical depth of the shell increases at first and then exhibits its peak at around $t\simeq 20$ s, followed by a steady decline toward $\tau=1$. 

\subsection{Some Important Time Scales}
We have assumed that the internal energy of the shell is dominated by radiation. 
This treatment implicitly assumes that the internal energy of the gas in the shell is efficiently converted to radiation. 
In the following, we check the consistency of the assumption. 

Some authors have already estimated the time scales of the ion-electron thermal equilibrium, photon production, and the Compton cooling, especially for non-relativistic cases \citep[e.g.,][]{2012ApJ...747L..17C}. 
We calculate these time scales in similar ways for our particular cases. 

\subsubsection{Condition of Ejecta at the Beginning of the Interaction}
At first, we estimate the energy density of the radiation field around the interface between the ejecta and the ambient medium at the beginning. 
Although we do not assume any specific process to produce the ejecta, they are supposed to be the stellar atmosphere accelerated and ejected by the passage of a strong shock wave. 
Thus, as usually expected in the context of supernova shock breakout, gas at the outermost layer of the ejecta is heated by the shock passage. 
For a compact star with the radius $\sim1R_\odot$ and the mass $\sim10\ M_\odot$, the temperature of the gas is expected to be of the order of $T_\mathrm{br}\sim 10^7\ \mathrm{K}$ after being swept by the shock and produce thermal photons at a similar temperature \citep{1999ApJ...510..379M}. 
Therefore, we can expect a radiation energy density of $u_\mathrm{ph}\simeq a_\mathrm{r}T_\mathrm{br}^4\sim 10^{14}\ \mathrm{erg}\ \mathrm{cm}^{-3}$, where $a_\mathrm{r}$ is the radiation constant, around the shell at the beginning of its dynamical evolution. 

We have assumed that the ejecta is cold when we formulate our semi-analytical model. 
The shocked gas actually has comparable amounts of the kinetic and internal energies after the passage of the shock wave. 
However, since the internal energy is soon lost by adiabatic expansion, the radiation pressure does not affect the dynamical evolution of the shocked gas so much. 
Thus, the assumption of the cold ejecta remains valid and our semi-analytical model still works well even when the initial temperature of $T_\mathrm{br}\sim 10^7\ \mathrm{K}$ is assumed. 

The number density $n_\mathrm{ph}$ of the photons at the outermost layer of the ejecta is estimated to be $a_\mathrm{r}T_\mathrm{br}^3/k_\mathrm{B}\sim 5\times 10^{22}\ \mathrm{cm}^{-3}$, where $k_\mathrm{B}$ is the Boltzmann constant. 
On the other hand, the number densities, $n_\mathrm{i}$ and $n_\mathrm{e}$, of ions and electrons, are of the order of $\rho_\mathrm{av}/m_\mathrm{u}\sim 10^{17}\ \mathrm{cm}^{-3}$, where $m_\mathrm{u}$ is the atomic mass unit, even when the density of the shell takes the maximum value of $\rho_\mathrm{av}\sim 10^{-7}\ \mathrm{g}\ \mathrm{cm}^{-3}$ (see, Figure \ref{fig:emission}). 
Thus, the photon number density dominates over the ion and electron number densities in the ejecta. 
This leads to an important consequence that the energy density of photons also dominates over those of ions and electrons when these particles efficiently exchange their energies and the temperature of these components takes the same value $T$, $n_\mathrm{ph}k_\mathrm{B}T\gg n_\mathrm{i}k_\mathrm{B}T,\ n_\mathrm{e}k_\mathrm{B}T$.

\subsubsection{Ion-Electron Equilibrium Time Scale}
Since the freely expanding ejecta primarily supply the shell with the energy and momentum through the reverse shock, we consider the shocked gas at the reverse shock front. 
Immediately after the passage of the shock front, ions and electrons are separately heated and eventually share the thermal energy. 
The ratio of the ion temperature $T_\mathrm{i}$ to the electron temperature $T_\mathrm{e}$ is at most $T_\mathrm{i}/T_\mathrm{e}\sim m_\mathrm{i}/m_\mathrm{e}$, where $m_\mathrm{i}$ and $m_\mathrm{e}$ are the masses of a single ion and electron. 
Thus, the thermal energy of ions is expected to dominate the internal energy of the gas immediately behind the reverse shock front. 
The time scale required to achieve the ion-electron thermal equilibrium depends on the microscopic process responsible for the energy exchange between ions and electrons. 

One of the well-known ways to establish the thermal equilibrium between ions and electrons is Coulomb collision. 
Ions heated by the shock passage repeatedly colliding with electrons to achieve the thermal balance. 
The time scale $t_\mathrm{ie}$ required for ions with the mass number $A_\mathrm{i}$ and the atomic number $Z_\mathrm{i}$ to lose a considerable fraction of its thermal energy is given by, 
\begin{equation}
t_\mathrm{ie}=\frac{3A_\mathrm{i}^2m_\mathrm{e}m_\mathrm{u}^2c^3}{8(2\pi)^{1/2}\rho Z_\mathrm{i}^3e^4\ln\Lambda}
\left(\frac{k_\mathrm{B}T_\mathrm{e}}{m_\mathrm{e}c^2}+\frac{k_\mathrm{B}T_\mathrm{i}}{A_\mathrm{i}m_\mathrm{u}c^2}\right)^{3/2},
\end{equation}
in cgs unit \citep[e.g.,][]{1962pfig.book.....S}, where $e$ is the elementary charge. 
Here, $\ln\Lambda$ is the so-called Coulomb logarithm and set to $\ln\Lambda=20$.  

\begin{figure}[tbp]
\begin{center}
\includegraphics[scale=0.55]{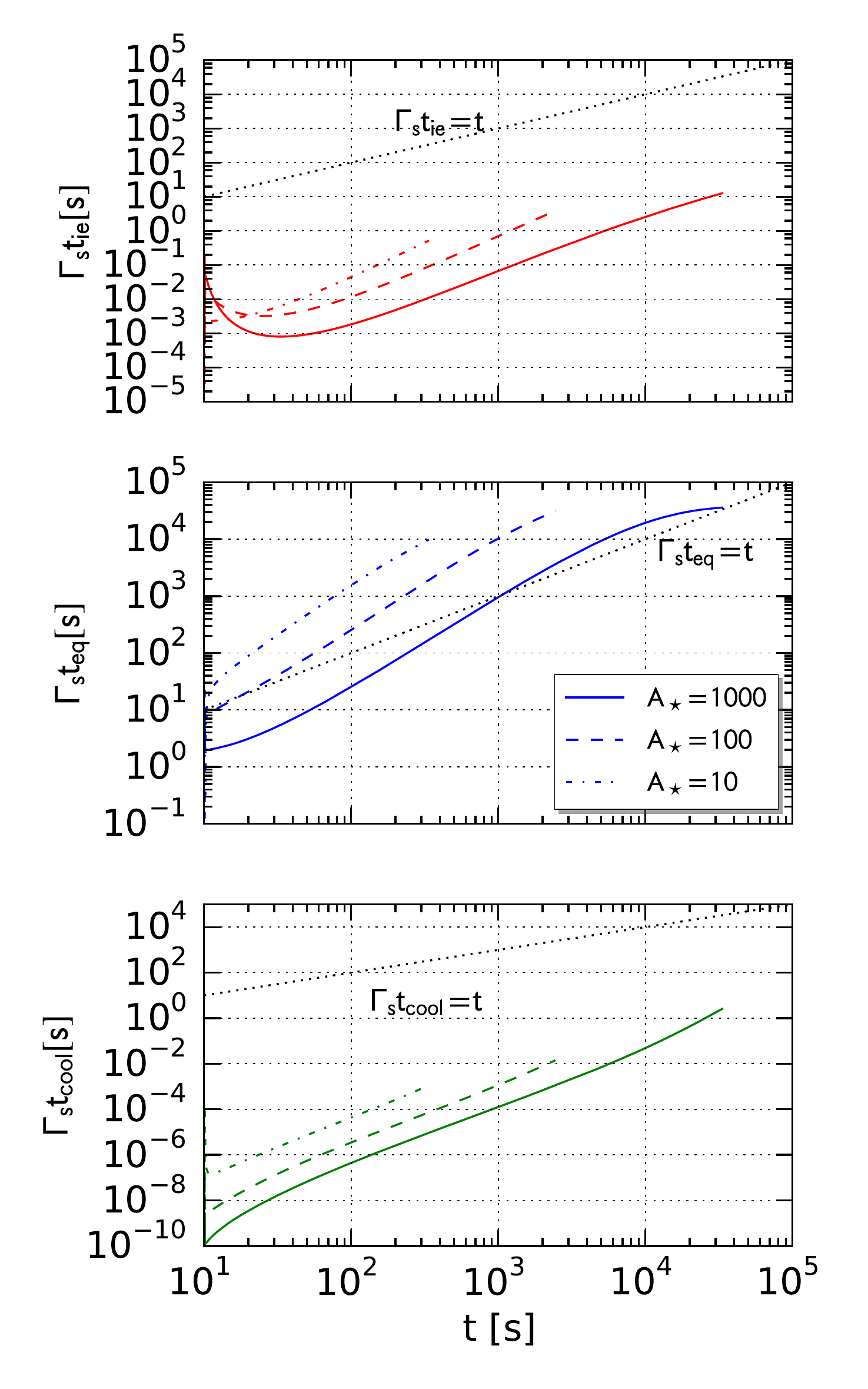}
\caption{Time scales for ion-electron thermal equilibrium $t_\mathrm{ie}$ (top), gas-radiation thermal equilibrium $t_\mathrm{eq}$ (middle), and Compton cooling $t_\mathrm{cool}$ (bottom) as functions of time $t$ for the same models as Figure \ref{fig:emission}.
The time scales are measured in the frame in which the center of the ejecta is at rest.}
\label{fig:time_scales}
\end{center}
\end{figure}

We calculate the time scale $t_\mathrm{ie}$ in the rest frame of the shell by assuming that the internal energy of the gas in the downstream of the reverse shock is dominated by that of ions, $T_\mathrm{i}\gg T_\mathrm{e}$, and using the following relation,
\begin{equation}
3p_\mathrm{rs}=\frac{\rho_\mathrm{rs} k_\mathrm{B}T_\mathrm{i}}{A_\mathrm{i}m_\mathrm{u}},
\end{equation}
to determine the ion temperature. 
Then, we obtain the corresponding time scale $\Gamma_\mathrm{s}t_\mathrm{ie}$ in the frame where the center of the ejecta is at rest. 
The top panel of Figure \ref{fig:time_scales} shows the calculated time scale $\Gamma t_\mathrm{ie}$ as a function of time $t$ for the same models as Figure \ref{fig:emission}. 
As shown in Figure \ref{fig:time_scales}, the equilibrium time scales are much shorter than the dynamical time $t$. 
We have also confirmed that the equilibrium time scales for models with smaller and larger kinetic energies, $E_\mathrm{rel,51}=0.01$, $0.1$, $1$, and $10$, are sufficiently short to achieve the ion-electron thermal equilibrium. 
Therefore, electrons in the downstream of the reverse shock are expected to possess a significant fraction of the internal energy of the shocked gas as well as ions. 

\subsubsection{Thermal Equilibrium Time Scale}
Thermal equilibrium between gas and radiation is achieved when processes absorbing and creating photons are quickly balanced. 
Since the shocked gas is hot and dilute, the most important radiative process to create photons is expected to be free-free emission. 
Therefore, we estimate the time scale $t_\mathrm{eq}$ required for free-free emission to produce radiation with an energy density $u_\mathrm{ph}$ comparable to the internal energy $u_\mathrm{int}$. 

Adopting the following frequency-integrated energy generation rate per unit volume per unit time for free-free emission,
\begin{equation}
\epsilon_\mathrm{ff}=5.5\times 10^{20}\rho^2T^{1/2}\frac{Z_\mathrm{i}^3}{A_\mathrm{i}^2}\ \mathrm{erg}\ \mathrm{cm}^{-3}\ \mathrm{s}^{-1}
,
\end{equation}
\citep[e.g.,][]{1979rpa..book.....R}, where the gaunt factor is neglected for simplicity, the time scale $t_\mathrm{eq}$ is estimated as follows,
\begin{equation}
t_\mathrm{eq}=\frac{u_\mathrm{int}'}{\epsilon_\mathrm{ff}},
\end{equation}
where $u_\mathrm{int}'$ is the internal energy density in the rest frame of the shell and is estimated by $u_\mathrm{int}'\sim u_\mathrm{int}/\Gamma_\mathrm{s}^2$. 
The temperature of the gas is calculated by assuming ion-electron thermal equilibrium. 
The middle panel of Figure \ref{fig:time_scales} shows the thermal equilibrium time scale $\Gamma_\mathrm{s}t_\mathrm{eq}$, which indicates that only models with significantly large ambient gas density $A_\star(\simeq 1000)$ can maintain the balance between absorption and creation of photons and produce radiation with the energy density $u_\mathrm{int}$.

\subsubsection{Compton Cooling Time Scale}
Electrons in the shocked gas can lose their energies via repeated Compton scattering. 
In the following, we estimate the time scale for a single electron swept by the reverse shock to lose a considerable fraction of the energy by scattering off photons. 

For the density and the internal energy density realized in this circumstance, the average energy of a single electron in the downstream of the reverse shock is much larger than the electron rest energy. 
In other words, the random motion of the electrons can be relativistic. 
This is a natural consequence because we can regard that ions and electrons are efficiently coupled and the shock is mildly relativistic. 
As seen in Figure \ref{fig:profile}, the ratio of the Lorentz factors of the pre-shocked and shocked ejecta is $\simeq 1.1$, which corresponds to a relative velocity of $v_\mathrm{rel}\simeq 0.4$. 
The kinetic energy per single ion of $\sim A_\mathrm{i}m_\mathrm{u}v_\mathrm{rel}^2$ is dissipated at the reverse shock and a considerable fraction of this energy is shared by ions and electrons in the downstream. 
Thus, the average Lorentz factor of the random motion of electrons in the shocked gas can be estimated from the condition, $\gamma_\mathrm{e}m_\mathrm{e}\simeq A_\mathrm{i}m_\mathrm{u}v_\mathrm{rel}^2/(Z_\mathrm{i}+1)$, which yields
\begin{equation}
\gamma_\mathrm{e}\simeq \frac{A_\mathrm{i}m_\mathrm{u}v_\mathrm{rel}^2}{(Z_\mathrm{i}+1)m_\mathrm{e}}=
2.8\times 10^2 \frac{A_\mathrm{i}}{Z_\mathrm{i}+1}\left(\frac{v_\mathrm{rel}}{0.4}\right)^2.
\end{equation}

The Compton cooling time $t_\mathrm{cool}$ of an electron with a Lorentz factor $\gamma_\mathrm{e}$ in a radiation field with an energy density $u_\mathrm{ph}$ is given by,
\begin{eqnarray}
t_\mathrm{cool}&=&\frac{3m_\mathrm{e}c}{4\sigma_\mathrm{T}u_\mathrm{ph}\gamma_\mathrm{e}}
\nonumber\\
&\simeq&4\times 10^{-6}\gamma_\mathrm{e}^{-1}\left(\frac{u_\mathrm{ph}}{10^{14}\ \mathrm{erg\ cm}^{-3}}\right)^{-1}
\ \mathrm{s}.
\label{eq:t_cool}
\end{eqnarray}
When the initially expected radiation energy density $u_\mathrm{ph}\sim 10^{14}\ \mathrm{erg}\ \mathrm{cm}^{-3}$ is assumed, the cooling time scale is much shorter than the dynamical time scale even for $\gamma_\mathrm{e}=1$. 
Thus, we can expect that electrons in the shell rapidly transfer their internal energies into radiation. 
Once the internal energy density of the gas is dominated by radiation, $u_\mathrm{int}\simeq u_\mathrm{ph}$, at an early epoch, the cooling time scale at later epochs can be estimated by substituting $u_\mathrm{int}'$, which is calculated by Equation (\ref{eq:u_int}), into $u_\mathrm{ph}$ in Equation (\ref{eq:t_cool}). 
As long as the condition $\Gamma_\mathrm{s}t_\mathrm{cool}\ll t$ is satisfied, the Compton cooling of electrons is efficient and thus the assumption $u_\mathrm{int}\simeq u_\mathrm{ph}$ is valid. 
The temporal evolution of the cooling time scales for the same models as in Figure \ref{fig:emission} are presented in the bottom panel of Figure \ref{fig:time_scales}, suggesting that $\Gamma_\mathrm{s}t_\mathrm{cool}\ll t$ is satisfied. 
Thus, we can assume that the energy exchange between electrons and photons is so efficient that $u_\mathrm{ph}\simeq u_\mathrm{int}$ is maintained even when the equilibrium between absorption and emission of photons is not achieved.

\subsection{Bolometric Light Curve}
\begin{figure*}[tbp]
\begin{center}
\includegraphics[scale=0.8]{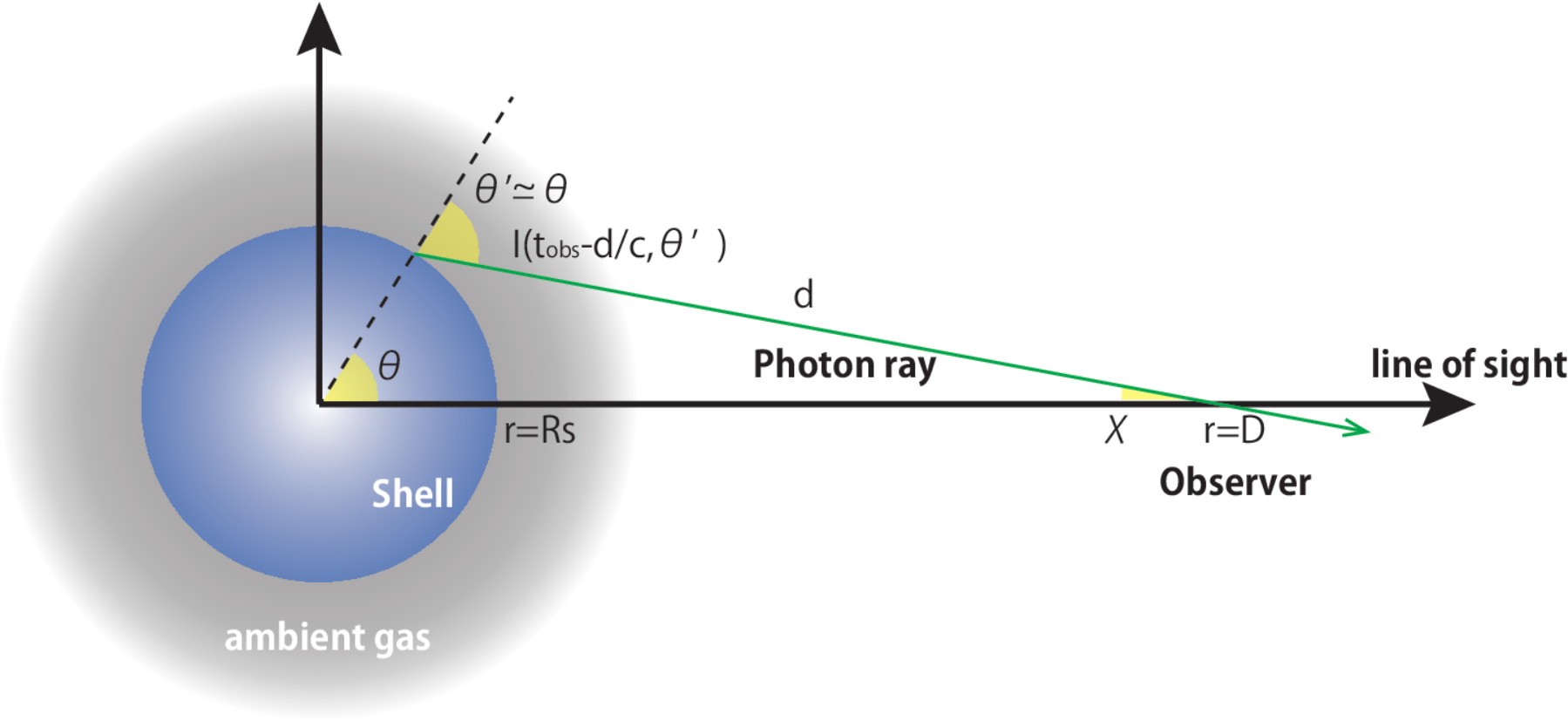}
\caption{Schematic view of the configuration considered in the light curve calculation.}
\label{fig:geometry}
\end{center}
\end{figure*}

We consider the bolometric light curve of the emission from the shell seen by a distant observer. 
Figure \ref{fig:geometry} schematically represents the geometry considered here. 
The observer sees the shell expanding and emitting photons. 
We consider a photon emitted from a part of the shell traveling at an angle $\theta$ with respect to the line of sight at time $t$. 
The intensity at the emitting point is given as a function of time $t$ and the angle $\theta'$ specifying the direction of the photon ray with respect to the radial direction. 
Assuming the distance $D$ from the center of the ejecta to the observer is much larger than the shell radius $R_\mathrm{s}$, the angle $\theta'$ between the photon ray and the radial direction is approximately equal to $\theta$, $\theta'\simeq \theta$. 
We regard that the photons reaching the observer at an angle $\chi$ with respect to the line of sight at time $t_\mathrm{obs}$. 
The photon travels at a distance $d$ given by
\begin{equation}
d=\sqrt{D^2-2DR_\mathrm{s}\cos\theta+R_\mathrm{s}^2}\simeq D-R_\mathrm{s}\cos\theta,
\end{equation}
before arriving at the observer, which leads to the following relation between the times, $t$ and $t_\mathrm{obs}$, of the emission and detection of the photon,
\begin{equation}
t=t_\mathrm{obs}-\frac{d}{c}\simeq t_\mathrm{obs}-\frac{D-R_\mathrm{s}\cos\theta}{c}.
\label{eq:t_obs}
\end{equation}
The angle $\chi$ is expressed in terms of $\theta$, $R_\mathrm{s}$, and $D$, in the following way,
\begin{equation}
\tan\chi=\frac{R_\mathrm{s}\sin\theta}{D-R_\mathrm{s}\cos\theta},
\end{equation}
which can be approximated as follows,
\begin{equation}
\chi \simeq \frac{R_\mathrm{s}}{D}\sin\theta.
\end{equation}
The intensity of the emission seen by the observer at the angle $\chi$ and time $t$ is identical with that at the emitting point at $\theta$ and $t_\mathrm{obs}-d/c$. 
Therefore, in the absence of an ambient medium, the flux at the observer is expressed as follows,
\begin{equation}
F(t_\mathrm{obs})
\simeq 
\frac{2\pi R^2}{D^2} \int_0^{\pi/2}I(t_\mathrm{obs}-d/c,\theta)\sin\theta \cos\theta d\theta.
\end{equation}

When the ambient medium ahead of the shell is optically thick, the energy of radiation diffusing out from the shell is at first deposited into the medium. 
How the deposited energy is transferred in the ambient medium before being detected by the observer depends on the optical depth $\tau_\mathrm{a}$ of the ambient medium measured from the emitting point specified by the radius $R_\mathrm{s}$ and the angle $\theta$ to the observer. 
Introducing a constant effective opacity $\kappa_\mathrm{eff}$, we calculate the optical depth $\tau_\mathrm{a}$ by integrating the density profile $\rho_\mathrm{a}(r)$ along the path. 
For a distance $D$ much larger than the shell radius $R_\mathrm{s}$, the photon travels almost parallel to the line of sight. 
Therefore, the optical depth can be approximated as follows,
\begin{equation}
\tau_\mathrm{a}(R_\mathrm{s},\theta)\simeq\frac{\kappa_\mathrm{eff}A}{R_\mathrm{s}}\int^\infty_0\frac{ds}{1+2s\cos\theta+s^2}.
\end{equation}
Electron scattering is again expected to be the dominant radiative process in the ambient medium, because the gas can be as hot as the outermost layer of the ejecta, whose temperature is expected to be $T\sim 10^7$ K, once it is illuminated by the emission from the layer. 
However, it is difficult to analytically determine when and what fraction of the radiation energy having deposited into the ambient medium escapes through the photosphere, since the transfer of the energy in the ambient medium includes scattering processes. 
Therefore, we adopt the following way to roughly estimate the bolometric luminosity at observer time $t_\mathrm{obs}$ and see how different values of $\kappa_\mathrm{eff}$ affect the bolometric light curve. 
A fraction $e^{-\tau_\mathrm{a}}$ of photons is expected to escape the ambient medium without being scattered by electrons. 
Thus, we only account for these freely escaping photons and calculate the flux of the emission in the following way,
\begin{eqnarray}
&&F(t_\mathrm{obs})\simeq\frac{2\pi R_\mathrm{s}^2}{D^2}
\nonumber\\
&&\times\int^{\pi/2}_0I(t_\mathrm{obs}-d/c,\theta)e^{-\tau_\mathrm{a}(R_\mathrm{s},\theta)}\sin\theta\cos\theta d\theta
.
\label{eq:Lbol}
\end{eqnarray}
The bolometric luminosity is given by
\begin{equation}
L_\mathrm{bol}(t_\mathrm{obs})=4\pi D^2F(t_\mathrm{obs}).
\end{equation}
When $\kappa_\mathrm{eff}=\kappa_\mathrm{es}$ is adopted, the luminosity evaluated above gives the lower limit on the bolometric luminosity, since we do not treat scattered photons, whose energy is $(1-e^{-\tau_\mathrm{a}})$ of the total amount. 
The scattered photons escape the ambient medium with a delay and make the emission brighter at later epochs. 
We also evaluate the bolometric luminosity with $\kappa_\mathrm{eff}=0$ and $0.1$ cm$^2$ g$^{-1}$. 
Setting $\kappa_\mathrm{eff}=0$ means that the deposited radiation energy directly seen by the observer and thus roughly gives the upper limit on the bolometric luminosity. 

Here we consider the angular distribution of the intensity. 
The radiation field is anisotropic due to the relativistic beaming effect, which makes the visible region from the observer only a part of the entire surface area of the shell. 
In order to see how the relativistic beaming affects the bolometric light curve, we employ the following simplified form of the intensity.  
When the shell is traveling at a velocity $\beta_\mathrm{s}$, outgoing photons emitted by the shell are confined in a cone with the direction cosine larger than the velocity, $\cos\theta\geq\beta_\mathrm{s}$. 
We simply assume that the intensity is uniform within the cone and otherwise vanishes,
\begin{equation}
I(t,\theta)=\left\{
\begin{array}{cl}
\frac{u_\mathrm{ph}}{\pi[(3+\beta_\mathrm{s}^2)\tau+2\beta_\mathrm{s}^2]}&\mathrm{for}\ \cos\theta\geq \beta_\mathrm{s},
\\
0&\mathrm{otherwise}.
\end{array}
\right.
\end{equation}
One can easily check that the flux multiplied by the surface area $4\pi R_\mathrm{s}^2$ of the shell is identical with the radiative energy loss rate, Equation (\ref{eq:dEdt2}),
\begin{eqnarray}
&&4\pi R_\mathrm{s}^2\int^{\pi/2}_02\pi I(t,\theta)\cos\theta\sin\theta d\theta
\nonumber\\
&&\hspace{4em}
=4\pi R_\mathrm{s}^2u_\mathrm{ph}\frac{1-\beta_\mathrm{s}^2}{(3+\beta_\mathrm{s}^2)\tau+2\beta_\mathrm{s}}.
\end{eqnarray}

In the calculation of bolometric light curves, we assume that the shell is at the origin, $R_\mathrm{s}=0$, at $t=0$. 
Then, the observer time corresponding to $t=0$ leads to $t_\mathrm{obs}=D/c$ from Equation (\ref{eq:t_obs}). 
However, this does not mean that the observation starts at $t_\mathrm{obs}=D/c$. 
When this event is observed by some instrument, the trigger of the observation would be delayed from $t_\mathrm{obs}=D/c$. 
Thus, we denote the trigger time of the observation by $t_\mathrm{obs,0}$ and plot light curves for different trigger times including $t_\mathrm{obs,0}=D/c$ to see how different trigger times affect the light curves. 

Finally, we should note that the shock curvature can effectively modify the bolometric light curve when the shock front is highly aspherical as suggested in the context of supernova shock breakout \citep{2010ApJ...717L.154S,2016ApJ...825...92S}. 
Although we assume a spherical shell in this study, highly aspherical blast waves may arise in stellar explosions producing trans-relativistic ejecta. 

\subsubsection{Emission Timescale}
At first, we consider the time scale of the emission. 
In the following consideration, we set $t_\mathrm{obs.0}=D/c$. 
When we focus on a part of the shell traveling parallel to the line of sight, $\theta=0$, the observer time is given by
\begin{equation}
t_\mathrm{obs}-t_\mathrm{obs,0}=t-\frac{R_\mathrm{s}}{c}.
\end{equation}
Thus, a shell traveling at a velocity close to the speed of light results in an emission time scale smaller than the time scale of the dynamical evolution of the shell. 
Assuming that the shell radius $R_\mathrm{tr}$ at the time $t=t_\mathrm{tr}$ is roughly given by the product $\beta_\mathrm{tr}t_\mathrm{tr}$ of the time and the velocity $\beta_\mathrm{tr}$ at the time, the emission time scale in the observer frame is smaller than $t_\mathrm{tr}$ by a factor of $(1-\beta_\mathrm{tr})$,
\begin{equation}
t_\mathrm{obs}-t_\mathrm{obs,0}=t_\mathrm{tr}(1-\beta_\mathrm{tr}).
\label{eq:emission_time1}
\end{equation}
Furthermore, for an ultra-relativistic speed, $\beta_\mathrm{tr}=1-1/(2\Gamma_\mathrm{tr}^2)$, the following familiar expression for relativistic flows can be obtained,
\begin{equation}
t_\mathrm{obs}-t_\mathrm{obs,0}\simeq \frac{t_\mathrm{tr}}{2\Gamma_\mathrm{tr}^2}.
\end{equation}

The shock curvature also affects the emission time scale. 
When photons are emitted from different parts of the shell at the same time, the distance between the nearest and the furthest points from the observer along the line of sight results in a delay in the arrival times, which is given by the light traveling time of the distance. 
This light traveling time effect is most significant when the shell radius is the largest at $t=t_\mathrm{tr}$. 
Since we take into account the relativistic beaming effect, the visible area is restricted to a part of the surface of the shell with $\cos\theta>\beta_\mathrm{tr}$. 
Thus, the distance between the center and the edge of the visible area along the line of sight leads to $R_\mathrm{tr}(1-\beta_\mathrm{tr})$. 
The corresponding light traveling time is given by,
\begin{equation}
\frac{R_\mathrm{tr}(1-\beta_\mathrm{tr})}{c}\simeq \beta_\mathrm{tr}t_\mathrm{tr}(1-\beta_\mathrm{tr})\simeq \frac{t_\mathrm{tr}}{2\Gamma_\mathrm{tr}^2},
\label{eq:emission_time2}
\end{equation}
where the last expression holds for ultra-relativistic velocities. 
Therefore, as long as the velocity $\beta_\mathrm{tr}$ when the shell becomes transparent to photons is close to the speed of light (not necessarily ultra-relativistic), Equations (\ref{eq:emission_time1}) and (\ref{eq:emission_time2}) give emission time scales of the same order. 

\subsubsection{Effective Opacity}
\begin{figure}[tbp]
\begin{center}
\includegraphics[scale=0.5]{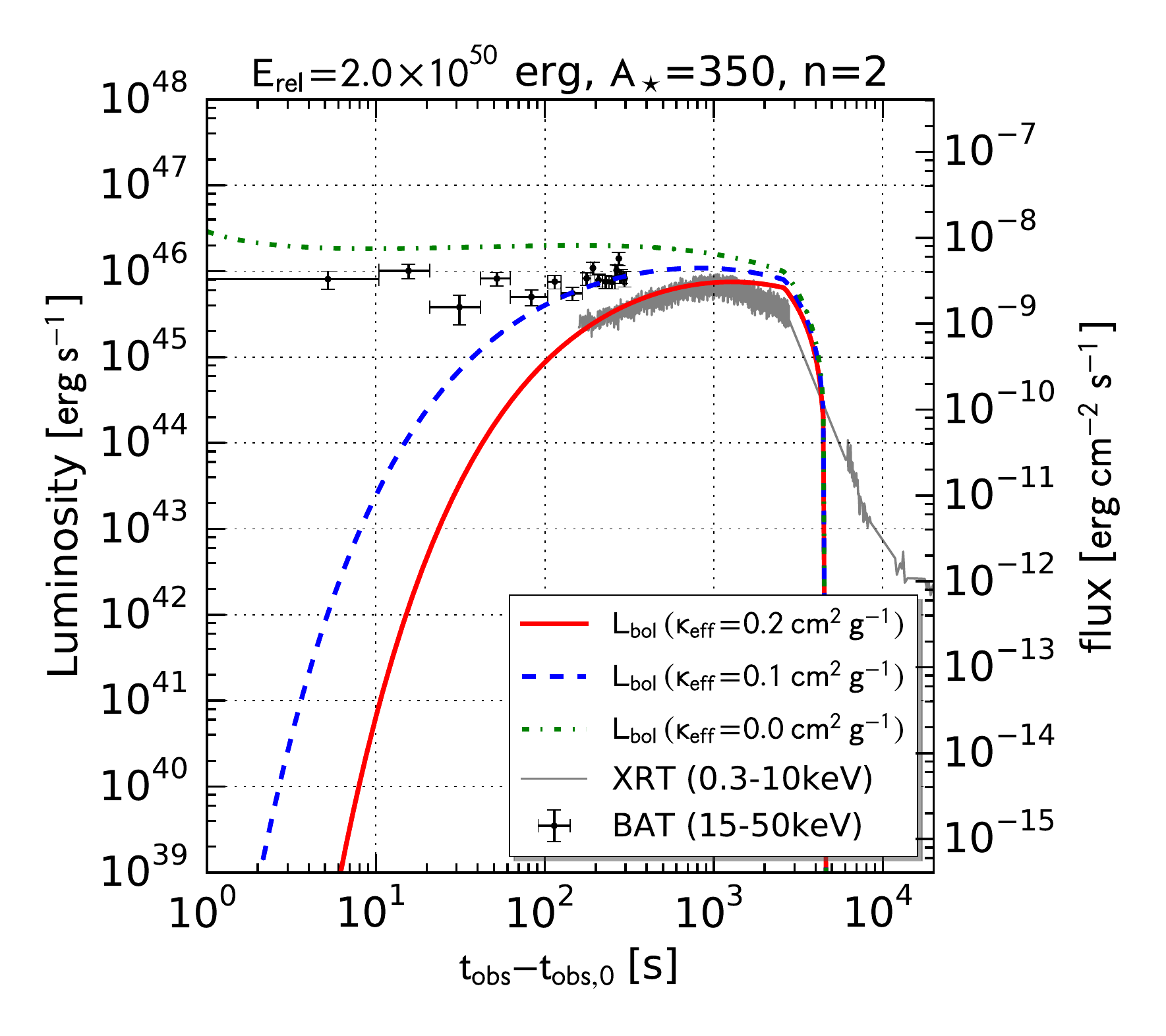}
\caption{Comparison of the bolometric light curve calculated by our model (thick lines) and the X-ray light curves of GRB 060218 obtained by {\it Swift} BAT ($15$-$50\ \mathrm{keV}$, circles with error bars) and XRT ($0.3-10\ \mathrm{keV}$, thin gray line). 
Thick solid (red), dashed (blue), and dash-dotted (green) lines show the bolometric light curves with different values of the effective optical depth, $\kappa_\mathrm{eff}=0.2$, $0.1$, and $0.0\ \mathrm{cm}^2\ \mathrm{g}^{-1}$ and a fixed $t_\mathrm{obs,0}=D/c$. 
The other free parameters of the model are set to $E_\mathrm{rel,51}=0.2$, $t_0=10$ s, $\Gamma_\mathrm{max}=5$, $A_\star=350$, and $n=2$. 
The distance to the source is assumed to be $D=143\ \mathrm{Mpc}$. }
\label{fig:lc}
\end{center}
\end{figure}

In Figure \ref{fig:lc}, we show the bolometric light curves calculated by our model with different values of $\kappa_\mathrm{eff}=0.2$, $0.1$, and $0.0$ cm$^{2}$ g$^{-1}$ and a fixed $t_\mathrm{obs,0}=D/c$, which are compared with the X-ray light curve of GRB 060218 in $0.3$-$10$ and $15$-$50$ keV obtained by {\it Swift} XRT and BAT. 
We use the data compiled and provided by UK Swift Science Data Centre\footnote{http://www.swift.ac.uk}. 
We note that our model only considers the bolometric luminosity of the emission and thus its spectrum may deviate from observations. 
The radiation energy emitted as X-ray photons is dominated by a small number of non-thermal photons, while thermal photons dominate the total photon number. 
We will further discuss the spectral features in Section \ref{sec:spec}.

The parameters of the model are set to $E_\mathrm{rel,51}=0.2$, $t_0=10$ s, $\Gamma_\mathrm{max}=5$, $A_\star=350$, and $n=2$. 
The emission is initially brighter for smaller values of the effective opacity $\kappa_\mathrm{eff}$ as expected. 
Although the discrepancy of the luminosities of the different models is considerably large at earlier epochs, $t_\mathrm{obs}<100\ \mathrm{s}$, it becomes smaller as the total optical depth of the ambient medium $\tau_\mathrm{a}$ becomes smaller. 
At later epochs $t_\mathrm{obs}>10^3$ s, the difference in the bolometric luminosity is within a factor of $3$.  
As a result, the overall duration and the average flux of the X-ray emission is well explained. 

\begin{figure}[tbp]
\begin{center}
\includegraphics[scale=0.5]{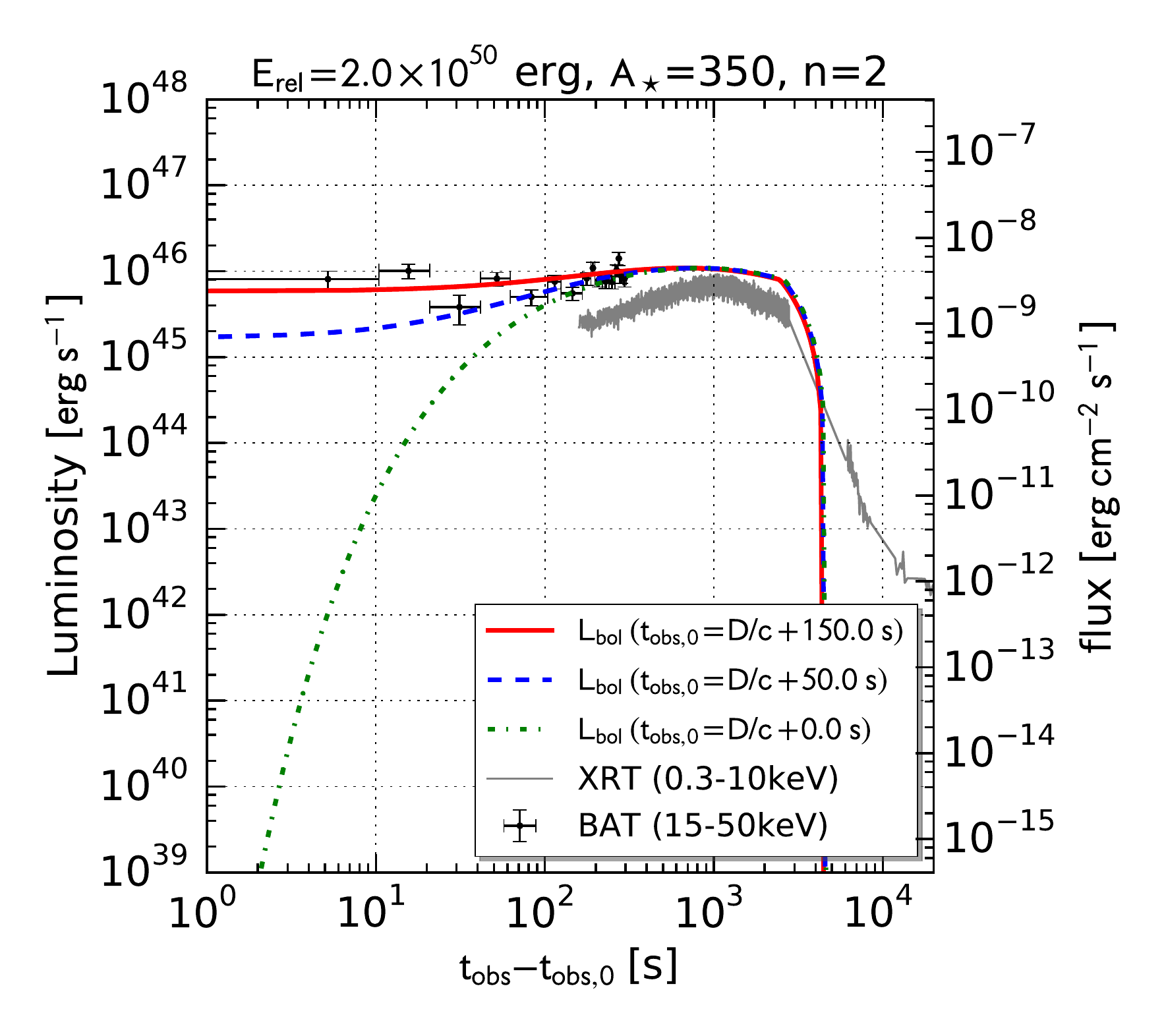}
\caption{Bolometric light curves with different values of $t_\mathrm{obs,0}$ and a fixed value of $\kappa_\mathrm{eff}=0.1\ \mathrm{cm}^2\ \mathrm{g}^{-1}$. 
The X-ray light curves of GRB 060218 is shown as in Figure \ref{fig:lc}. 
Thick solid (red), dashed (blue), and dash-dotted (green) lines correspond to $t_\mathrm{obs,0}=D/c+150$ s, $D/c+50$ s, and $D/c$. 
The same free parameters as the models shown in Figure \ref{fig:lc} are used. }
\label{fig:lc_dt}
\end{center}
\end{figure}

\subsubsection{Trigger Time of Observations}
The light curve models shown in Figure \ref{fig:lc} exhibit a clear offset between the rise of the light curve and $t_\mathrm{obs}=t_\mathrm{obs,0}$, depending on the adopted values of $\kappa_\mathrm{eff}$. 
This suggests that the trigger time $t_\mathrm{obs,0}$ of the observation significantly affects the light curve.  
In Figure \ref{fig:lc_dt}, we plot the light curves of models with different trigger times $t_\mathrm{obs,0}$ and a fixed $\kappa_\mathrm{eff}=0.1$ cm$^2$ g$^{-1}$. 
The trigger time is set to $t_\mathrm{obs,0}=D/c$, $D/c+50$ s, and $D/c+150$ s. 
The results clearly demonstrate that the earlier part of the light curve is significantly affected by the different values of $t_\mathrm{obs,0}$, while the later part remains unchanged. 
Especially, the light curve with $t_\mathrm{obs,0}=D/c+150$ s matches the observed one. 
Thus, when the emission is observed after its bolometric luminosity reaches around $L_\mathrm{bol}\simeq 6\times 10^{45}\ \mathrm{erg}\ \mathrm{s}^{-1}$, the observed light curve can be well reproduced. 

\subsubsection{Termination of the Emission}
In our semi-analytical model, the energy deposition from the shell via radiative diffusion is terminated when the optical depth of the shell $\tau$ becomes smaller than unity, $\tau<1$.
This termination happens at $t_\mathrm{obs}\simeq 3\times 10^3$ s for the model shown in Figures \ref{fig:lc} and \ref{fig:lc_dt}. 
Interestingly, the X-ray light curve of GRB 060218 also exhibits a sharp drop in luminosity at a similar epoch. 
In our model, this sudden decline in the luminosity is interpreted as the transition of the shell from optically thick to thin, which leads to a significantly small number of photons coupled to electrons.

\subsection{$E_\mathrm{rad}$-$T_{90}$ Diagram}
Results shown in Figures \ref{fig:lc} and \ref{fig:lc_dt} indicate that the bolometric luminosity at later epochs ($t_\mathrm{obs}>10^3$ s) less suffers from the uncertainties mentioned above, the radiative transfer in the ambient medium and the trigger time. 
This suggests that the duration of the bright emission, the luminosity at later epochs, and the total radiated energy, which is roughly given by the duration and the late-time luminosity, are appropriate quantities for a fair comparison between our models and observations. 
Observations of individual GRBs provide the duration and isotropic gamma-ray energy. 
Thus, we define the following quantities, which can easily be calculated from our light curves, and compare them with duration and isotropic gamma-ray energy of several nearby GRBs. 

Integrating the bolometric luminosity $L_\mathrm{bol}$ with respect to the observer time $t_\mathrm{obs}$, we obtain the total radiated energy $E_\mathrm{rad}$,
\begin{equation}
E_\mathrm{rad}=\int_0^\mathrm{t_\mathrm{obs,max}}L_\mathrm{obs}(t_\mathrm{obs})dt_\mathrm{obs},
\end{equation}
with $t_\mathrm{obs}=10^5$ s. 
Furthermore, we define a time scale $T_{90}$ at which 90 $\%$ of the total radiation energy has been deposited\footnote{Note that the definition of $T_{90}$ is different from that used in observations. In this paper, the latter is denoted by $T_\mathrm{90,obs}$.},
\begin{equation}
\int_0^\mathrm{T_{90}}L_\mathrm{obs}(t_\mathrm{obs})dt_\mathrm{obs}=0.9E_\mathrm{rad}.
\end{equation}

\begin{table*}
\begin{center}
  \caption{Properties of selected GRB-SNe}
\begin{tabular}{llrrrr}
\hline\hline
GRB&SN&$T_\mathrm{90,obs}$ [s]&$E_\mathrm{\gamma,iso} $[erg]&$E_\mathrm{peak}$ [keV]&Reference\\
\hline
GRB 980425&SN 1998bw&$34.9\pm 3.8$&$9\times 10^{47}$&$122\pm17$&1\\
GRB 030329&SN 2003dh&$22.9$&$1.3\times 10^{52}$&$70\pm 2$&1\\
GRB 031203&SN 2003lw&$37.0\pm1.7$&$1.3\times 10^{50}$&$>71$&1\\
GRB 060218&SN 2006aj&$2100\pm100$&$4\times 10^{49}$&$4.7\pm 1.2$&1\\
GRB 100316D&SN 2010bh&$1300$&$6\times 10^{49}$&$18^{+3}_{-2}$&2,3\\
GRB 120422A&SN 2012vz&$5.35\pm1.4$&$4.4\times 10^{49}$&$53$&4,5,6\\
\hline\hline
\end{tabular}
\tablerefs{(1) \cite{2007ApJ...654..385K}; (2) \cite{2011MNRAS.411.2792S}; (3) \cite{2012grbu.book..169H}; (4) \cite{GCN13246}; (5) \cite{GCN13257}; (6) \cite{2012ApJ...756..190Z}}
 \label{table:GRB}
\end{center}
\end{table*}

We calculate these two quantities from models with various sets of the parameters $(E_\mathrm{rel},A_\star,n)$, while the maximum Lorentz factor is fixed $\Gamma_\mathrm{max}=5$. 
In each panel of Figure \ref{fig:diagram}, the relations between the radiated energy $E_\mathrm{rad}$ and the time scale $T_\mathrm{90}$ for various sets of the parameters are plotted. 
Solid lines represent the relations of these two quantities obtained by increasing the parameter $A_\star$ from $A_{\star}=1$ to $A_\star=1000$ for $E_\mathrm{rel,51}=0.01$, $0.1$, $1$, $10$. 
On the other hands, thin dashed lines represent the relations obtained by increasing the kinetic energy $E_\mathrm{rel,51}$ from $0.01$ to $10$ for $A_{\star}=1$, $10$, $100$, and $1000$. 

\subsubsection{Nearby Less-energetic GRBs}
\begin{figure*}[tbp]
\begin{center}
\includegraphics[scale=0.6]{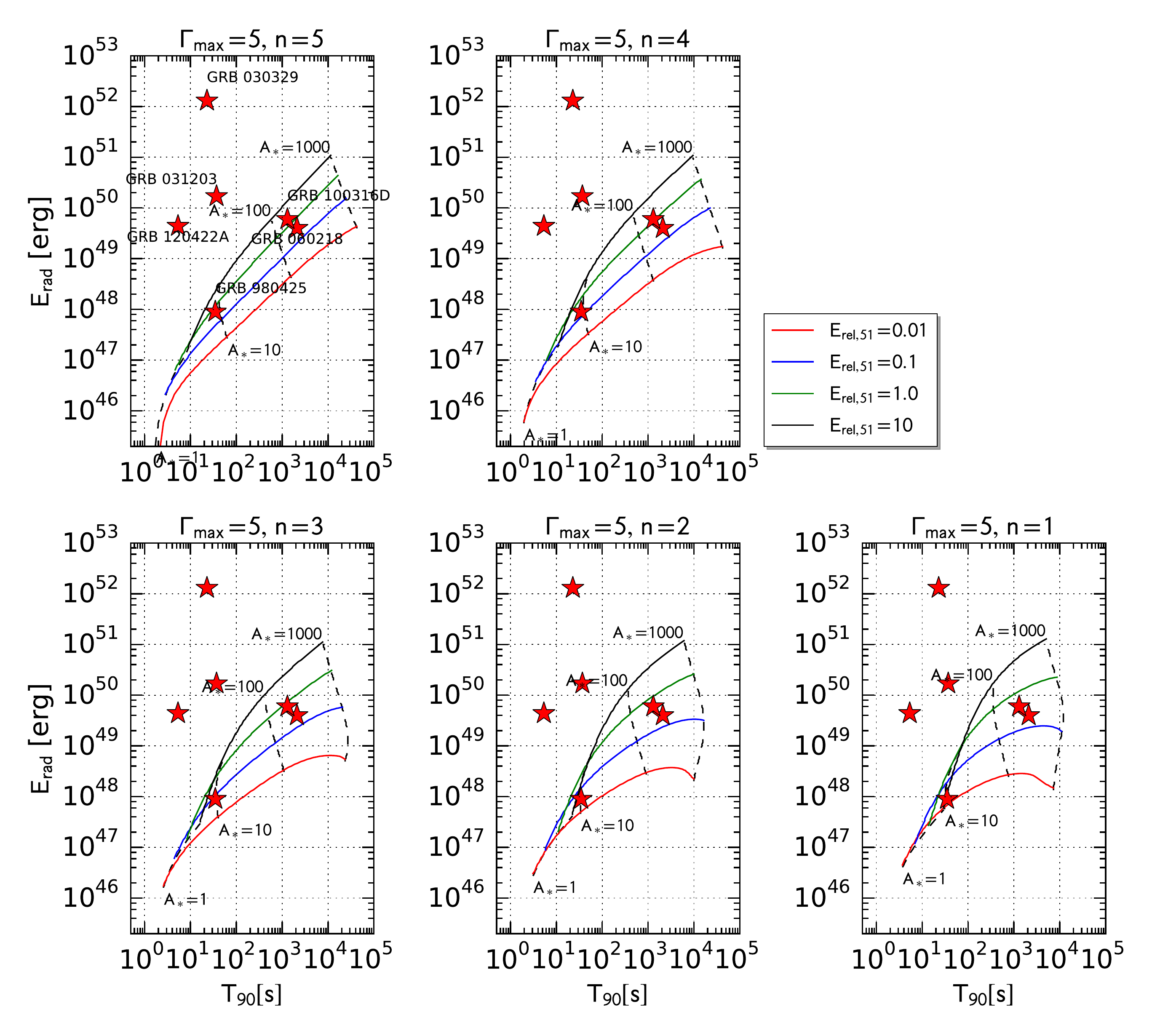}
\caption{Diagrams showing the total radiated energy $E_\mathrm{rad}$ and the duration $T_{90}$ of the interaction powered emission for various sets of the free parameters. 
Solid lines show relations of $E_\mathrm{rad}$ and $T_\mathrm{90}$ obtained by increasing the parameter $A_\star$ from $A_{\star}=1$ to $A_{\star}=1000$ for $E_\mathrm{rel,51}=0.01$ (red), $0.1$ (blue), $1.0$ (green), and $10$ (black). 
Dashed lines show those obtained by increasing the energy $E_\mathrm{rel}$ from $E_\mathrm{rel,51}=0.1$ to $E_\mathrm{rel,51}=10$ for $A_{\star}=1$, $10$, $100$, and $1000$. 
The duration $T_\mathrm{90,obs}$ and isotropic energy $E_\mathrm{\gamma,iso}$ of the selected GRBs, which are summarized in Table \ref{table:GRB}, are also plotted.  }
\label{fig:diagram}
\end{center}
\end{figure*}

We compare the resultant quantities with the duration $T_\mathrm{90,obs}$ and the isotropic gamma-ray energy $E_\mathrm{\gamma,iso}$ of nearby GRBs associated with SNe, some of which are less energetic than normal GRBs found at cosmological distances. 
The properties of the gamma-ray emission of the selected GRBs are summarized in Table \ref{table:GRB}. 
These quantities are plotted on the panels in Figure \ref{fig:diagram}. 
Out of these samples, GRBs 980425, 060218, and 100316D are often classified as LLGRBs due to their small gamma-ray isotropic energies. 
Furthermore, GRB 060218 and 100316D exhibit quite similar X-ray light curves, characterized by 1000 sec-long duration and soft X-ray spectrum. 
On the other hand, GRB 030329 shows similar properties, such as the isotropic gamma-ray energy and the spectral peak energy, to normal GRBs rather than LLGRBs. 
A more recently detected GRB 120422A shows gamma-ray emission with a relatively short duration, $T_{90}=5.35\pm 1.4$ s, followed by rapidly fading X-ray emission. 
It has been suggested that it might constitute the intermediate class between normal GRBs and LLGRBs \citep{2014A&A...566A.102S}. 

\subsubsection{Interaction-powered Emission}
From the $E_\mathrm{rad}$-$T_\mathrm{90}$ diagrams in Figure \ref{fig:diagram}, LLGRBs 980425, 060218, and 100316D, are found to be well explained by the interaction-powered emission from spherical ejecta. 
GRBs 060218 and 100318D are best explained by interaction-powered emission from relativistic ejecta with a kinetic energy of $E_\mathrm{rel,51}=0.1$-$1.0$ and a CSM with $A_\star\simeq 300$-$400$. 
The large value of the parameter $A_\star$ required to explain their long-lasting X-ray emission implies a mass-loss rate close to $(3-4)\times 10^{-3}\ M_\odot$ yr$^{-1}$ when a typical wind velocity of Wolf-Rayet stars, $10^3$ km s$^{-1}$, is adopted. 
The implied large amount of CSM has also been pointed out by several authors since the discovery of GRB 060218 \citep[e.g.,][]{2006Natur.442.1008C,2007ApJ...667..351W}. 
GRB 980425 is reproduced by models with moderate values of the parameter $A_\star$ and an even small kinetic energy, $E_\mathrm{rel,51}=0.01$-$0.1$, is sufficient to account for the isotropic gamma-ray energy of this relatively less energetic event. 
For larger values of the exponent $n$, a larger kinetic energy is required for GRB 980425, since a larger $n$ means that a smaller fraction of the kinetic energy is distributed throughout the outer layers of the ejecta, resulting in emission with longer rise times. 

\subsubsection{Central Engine Driven Bursts}
Several GRBs are located above the lines corresponding to models with the largest kinetic energy $E_\mathrm{rel,51}=10$ in all panels of Figure \ref{fig:diagram}. 
In other words, they cannot be explained by the interaction-powered emission. 
Assuming ejecta with a kinetic energy larger than $E_\mathrm{rel,51}=10$ is less likely, because the value $E_\mathrm{rel,51}=10$ is already comparable to the kinetic energy of particularly energetic SNe, i.e., hypernovae, and a considerable fraction of the total kinetic energy of SN ejecta is usually distributed in slower non-relativistic ejecta rather than outer layers. 
Thus, it seems unfeasible to distribute even more kinetic energy throughout the faster part of the ejecta while the spherical symmetry is kept. 
These bursts require relativistic jets powered by central engines

\subsection{Spectral Features}\label{sec:spec}
Our model predicts bolometric properties of the emission from the shocked gas and provides little information on the spectrum. 
Nevertheless, it is worth discussing the following points regarding the expected spectrum. 

The discussion on the thermal equilibrium time scale implies that the balance between absorption and emission of photons via free-free process is achieved at early stages of the dynamical evolution of the shell, $t\simeq20$ s and $10^3$ s for $A_\star =100$ and $1000$ (see Figure \ref{fig:time_scales}). 
In particular, for the model shown in Figures \ref{fig:lc} and \ref{fig:lc_dt}, the decoupling occurs at $t\simeq 150$ s. 
The reverse shock pressure at the time is $p_\mathrm{rs}\simeq 10^{11}\ \mathrm{erg}\ \mathrm{cm}^{-3}$, leading to an equilibrium temperature of $T_\mathrm{eq}\simeq 2.5\times 10^6$ K. 
Thus, these photons are expected to show a temperature of  a few $10^6$ K or equivalently an average energy of $3k_\mathrm{B}T_\mathrm{eq}\sim 0.6$ keV at the decoupling and contribute to the emission as a thermal component. 
In fact, a thermal component has been found in the X-ray spectra of GRB 060218 \citep{2006Natur.442.1008C} and 100318D \citep{2011MNRAS.411.2792S}. 
The presence of a thermal component with temperature of $k_\mathrm{B}T_\mathrm{BB}\sim 0.1-0.2$ keV, relatively low spectral peak energy ($E_\mathrm{peak}=4.7$ and 18 keV, respectively), and their peculiarly long X-ray emission may indicate that their emission originates from the interaction between a dense CSM and mildly relativistic ejecta. 
The contribution of the thermal component to the total X-ray flux is only $\sim 10$ \% and $2$-$4$ \% for GRB 060218 and 100318D. 
However, the low temperature, $0.1$-$0.2$ keV, compared with the spectral peak energy means that thermal photons dominate the total photon number. 

Our model suggests that thermal photons with a temperature of $\sim 0.1$-$0.2$ keV predominantly exist until the decoupling, which reasonably agrees with the observations. 
Since the energy injection through the reverse shock continues even after the decoupling, the freshly injected energy would play a role in powering a small fraction of photons remaining in the shocked region and producing a non-thermal component. 
For the model shown in Figures \ref{fig:lc} and \ref{fig:lc_dt}, the internal energy of the shell at the decoupling $t\simeq 150$ s is $E_\mathrm{int,dec}\simeq 10^{49}$ erg, while the value at $t=t_\mathrm{tr}$ is $E_\mathrm{int,tr}\simeq 2.6\times 10^{49}$ erg. 
Thus, an additional energy of the order of $10^{49}$ erg can be used to further power the photons. 
Because of the freeze-out of the photon number, when this additional energy is simply used to increase the average energy of the whole remaining photons, the increase in the photon energy, which is given by $E_\mathrm{int,tr}/E_\mathrm{int,dec}$, is only a factor of a few. 
The increased average photon energy is $\sim 1$ keV, which does not explain the spectral peak energies of LLGRBs. 
This is the reason why some mechanism decoupling a small fraction of photons from the thermal equilibrium should play a role. 
For example, when a few to 10 percent of the remaining photons could predominantly gain the additional energy, an energy larger than the average photon energy by a factor of $10$-$30$, $\sim 6$-$10$ keV, would be achieved. 
In fact, this simple consideration overestimates the contribution of the thermal photons to the total radiated energy, $E_\mathrm{int.dec}/E_\mathrm{int,tr}\sim 38\%$. 
Thus, a more sophisticated model taking into account the process producing non-thermal photons should be developed to explain observations of LLGRBs in a self-consistent way. 
However, revealing the mechanism responsible for producing the non-thermal component is beyond the reach of this study, where a simplified one-zone model is used to describe the shocked region. 


There is also the possibility that the spherical shell is not responsible for producing non-thermal photons and the non-thermal component can instead arise from a weak jet rather than the shell \citep[see,][for a recent discussion]{2016MNRAS.460.1680I}. 
In other words, the combination of non-thermal emission from a jet and thermal emission from the associated quasi-spherical ejecta, such as a cocoon, accounts for the emission from LLGRBs. 
In this case, just up to $\sim 10$ \% of the total X-ray flux should be explained by the ejecta-ambient interaction and make the required kinetic energy of the ejecta smaller by an order of magnitude. 

\section{CONCLUSIONS AND DISCUSSIONS}\label{sec:conclusions}
In this study, we have developed a semi-analytical model for the hydrodynamical evolution of freely expanding trans-relativistic ejecta interacting with an ambient medium whose density is inversely proportional to the square of the radius. 
Our model successfully reproduces results of numerical simulations. 
Furthermore, we have presented an emission model in which photons are diffusing out from the geometrically thin and optically thick shell resulting from the hydrodynamical interaction. 
Then, we have examined the possibility that the CSM interaction contributes to X-ray and gamma-ray emission from GRBs associated with SNe, including several events classified as LLGRBs. 
Our results give a threshold on the $E_\mathrm{rad}$-$T_{90}$ plane, above which gamma-ray emission of any GRB cannot be explained by the interaction-powered emission from mildly relativistic, spherical ejecta. 
The CSM interaction cannot explain GRBs releasing a large amount of gamma-ray energy during short duration, such as GRB 030329, 031203, and 120422A, which suggests that they require a certain central engine activity to deposit a significant amount of energy into a small solid angle. 
This result agrees with \cite{2012ApJ...756..190Z}, who claimed that the time-averaged luminosity of the gamma-ray emission could distinguish central-engine powered GRBs from LLGRBs. 

\subsection{Presence of Dense CSM}
Our results suggest that the circum-burst environment would be a key to distinguishing GRBs characterized by long-lasting soft X-ray emission from normal GRBs. 
Recently, \cite{2015ApJ...805..159M} studied X-ray emission from GRBs found in the nearby universe and found that some GRBs exhibit the following properties: 
(1) long-lasting gamma-ray emission with $T_\mathrm{90,obs}>1000$ s, 
(2) a large absorption column density by neutral hydrogen, and 
(3) late-time super soft X-ray emission. 
They pointed out that these events were possibly explosions of stars surrounded by dense materials having been lost from the stellar surface as a stellar wind, although they could not exclude another possibility that these events originate from unusual progenitor systems.

It is found that a considerably dense CSM, mass-loss rates larger than $10^{-3}$ $M_\odot$ yr$^{-1}$ for a wind velocity of $10^3$ km s$^{-1}$, is required for GRB 060218 and 100316D. 
How such dense CSM is produced during the evolution toward the core-collapse is still unclear. 
However, spectroscopic observations of SNe at very early stages may shed light on the origin. 
Recent observations of type IIb SN 2013cu \citep{2014Natur.509..471G} at the very beginning of its evolution suggested that it was the explosion of a Wolf-Rayet star with an extremely dense CSM or stellar envelope attached. 
From the luminosity of the H$\alpha$ line emission found in the early optical spectrum of SN 2013cu, they argued the presence of a strong stellar wind with a mass-loss rate of the order of $10^{-2}\ M_\odot$ yr$^{-1}$ and a wind velocity of the order of $10^3$ km s$^{-1}$ in the vicinity of the progenitor star. 
More detailed spectral modeling later found that the wind velocity appeared to be smaller than the value obtained by  \cite{2014Natur.509..471G}, but the mass-loss rate was similar \citep{2014A&A...572L..11G,2016MNRAS.455..112G}. 
Furthermore, recent spectroscopic observations of SNe found by the Palomar Transient Factory suggest that a non-negligible fraction of the SNe exhibits spectral signatures similar to SN 2013cu \citep{2016ApJ...818....3K} in their early optical spectra. 
If a highly energetic explosion producing relativistic ejecta occurs in a similar environment, the ejecta-CSM interaction would give rise to bright X-ray or gamma-ray emission like GRBs 060218 and 100316D. 

On the other hand, radio observations and light curve modelings of GRBs associated with SNe imply relatively dilute circum-burst environments. 
For example, \cite{2013ApJ...778...18M} carried out radio and X-ray observations of GRB 100316D and estimated the mass-loss rate of $\sim 10^{-5}$ $M_\odot$ yr$^{-1}$ for a wind velocity of $10^3$ km s$^{-1}$. 
This discrepancy could be resolved by introducing a non-steady mass losing process. 
In other words, radio observations would probe the CSM at more distant regions from the explosion site than the CSM predominantly dissipating the kinetic energy of the ejecta. 
The geometrically thin shell becoming transparent at $t_\mathrm{tr}\sim10^3$ s is located at a distance of $ct_\mathrm{tr}\sim 3\times 10^{13}$ cm, where materials ejected a few days before the explosion should be present for a wind velocity of $10^3\ \mathrm{km}\ \mathrm{s}^{-1}$. 
On the other hand, radio observations are usually carried out 10-100 days after the discovery, when relativistic ejecta traveling at speeds close to the speed of light have reached at a distance of $3\times 10^{16-17}$ cm. 
The corresponding traveling time scale of the wind at a velocity $10^3\ \mathrm{km}\ \mathrm{s}^{-1}$ is $10$-$100$ yrs. 
Thus, the drastic change in the mass-loss history of the progenitor, if it existed, must have occurred $10$-$100$ yrs before the explosion.

Radio observations of LLGRBs also suggest that the radio-emitting ejecta still have sub-relativistic velocities, $\Gamma\beta\sim 1$, and the kinetic energy of the ejecta is comparable to the isotropic energy of the gamma-ray emission \citep[e.g.,][]{2013ApJ...778...18M}. 
If the shell resulting from the ejecta-CSM interaction continued to decelerate in a dense CSM, the shell velocity would be non-relativistic and most of the kinetic energy would be lost at later stages of the dynamical evolution. 
Thus, the termination of the dense CSM in a distant region from the explosion site is also required to explain the dynamics of radio emitting ejecta. 


\subsection{Kinetic Energy Distribution of Trans-relativistic Ejecta}
The total radiated energy and the duration of the emission predicted by our emission model do not strongly depend on the gradient of the assumed kinetic energy distribution, which is characterized by the power-law exponent $n$, especially for ambient media with high densities. 
However, the temporal evolution of the spectrum of the emission must depend on $n$. 
Smaller values of $n$ correspond to ejecta with larger kinetic energy at the outermost layer. 
Ejecta traveling at higher Lorentz factor are expected to emit harder photons due to the Lorentz boost. 
Therefore, when ejecta with a kinetic energy distribution with a smaller $n$ are responsible for the X-ray and gamma-ray emission, the spectrum is dominated by hard photons, while a softer spectrum is expected for ejecta with a steep kinetic energy distribution. 
The difference in the hardness of observed spectra must be a key to investigating what kind of kinetic energy distribution is realized in LLGRBs. 

How the assumed kinetic energy distribution is related to the hardness of the resultant emission should be quantitatively studied by performing hydrodynamic simulations with multi-frequency radiative transfer, which would also clarify whether a thermal component is really present in the spectrum of the emission and how repeated Compton scatterings shape the spectrum. 

\acknowledgments
We appreciate the anonymous referee for his/her constructive comments, which greatly helped us to improve the manuscript. 
Numerical calculations were in part carried out on the XC30 system at the Center for Computational Astrophysics, National Astronomical Observatory of Japan.
AS is supported by Grant-in-Aid for JSPS Research Fellow (26$\cdot$10618). 
The work by KM is partly supported by Japan Society for the Promotion of Science (JSPS) KAKENHI Grant 26800100 and by World Premier International Research Center Initiative (WPI Initiative), MEXT, Japan.

\appendix
\section{Diffusion Approximation in Relativistic Regime}\label{sec:diffusion}
In this section, we describe the method to evaluate the energy loss rate through radiative diffusion from a geometrically thin and optically thick shell moving at a relativistic speed. 
In the following, we consider two inertial frames, the laboratory frame, in which the shell is traveling at a velocity $\beta_\mathrm{s}$, and the comoving frame, in which the shell is at rest. 
We denote quantities in the laboratory and comoving frames by letters with and without a prime, e.g., $Q$ and $Q'$. 

Since the radiative energy loss rate does not change under the Lorentz transformation, $(dE/dt)_\mathrm{rad}=(dE'/dt')_\mathrm{rad}$, we derive the rate in the comoving frame at first. 
The transfer equation for the frequency-integrated intensity $I'$ with spherical symmetry is given by,
\begin{equation}
\frac{\partial I'}{\partial t'}+\mu'\frac{\partial I'}{\partial r'}+\frac{1-{\mu'}^2}{r'}\frac{\partial I'}{\partial \mu'}=\rho'\kappa'(J'-I'),
\label{eq:transfer}
\end{equation}
where $t'$ and $r'$ are the time and the radial coordinate and $\mu'$ is the direction cosine. 
For simplicity, we assume isotropic scattering and $\kappa'$ gives the scattering opacity. 
In the diffusion approximation, we only consider the zeroth and first angular moments of the intensity. 
Thus, the intensity is expressed as follows,
\begin{equation}
I'=J'+\mu' H.
\label{eq:I'}
\end{equation}
The zeroth and first angular moments of the transfer equation (\ref{eq:transfer}) with Equation (\ref{eq:I'}) lead to
\begin{equation}
\frac{\partial J'}{\partial t'}+\frac{1}{3r'^2}\frac{\partial (r'^2H')}{\partial r'}=0,
\label{eq:dJdt}
\end{equation}
and
\begin{equation}
\frac{\partial H'}{\partial t'}+\frac{\partial J'}{\partial r'}=\rho'\kappa'H'.
\label{eq:dHdt}
\end{equation}
As usually done in the diffusion approximation \cite[e.g.,][]{1979rpa..book.....R}, we assume that the time derivative of the first moment $\partial H'/\partial t'$ is much smaller than the other terms in the above equation. 
Thus, Equation (\ref{eq:dHdt}) yield
\begin{equation}
H'=\frac{1}{\rho'\kappa'}\frac{\partial J'}{\partial r'}.
\label{eq:H'}
\end{equation}

Here we consider a spherical shell with the outer radius $R'$ and width $L'$. 
Furthermore, we assume that the radiation energy of the shell is lost from $r'=R$ via radiative diffusion. 
We multiply Equation (\ref{eq:dJdt}) by $4\pi r'^2$ and integrate the product with respect to the radius from $r'=R'-L'$ to $r'=R'$. 
Further multiplying the result with $4\pi$, which corresponds to the integration over the solid angle, one obtains the rate of the radiative energy loss from the outer radius,
\begin{equation}
\frac{dE'}{dt'}=-\frac{16\pi^2 R'^2}{3}H'|_{r'=R'}.
\end{equation}
When the width of the shell is negligibly small as assumed in this paper, Equation (\ref{eq:H'}) can be approximated as follows,
\begin{equation}
H'=\frac{J'}{\rho'\kappa'L'}=\frac{J'}{\tau}.
\label{eq:H2}
\end{equation}
Here, $\tau$ is the optical depth of the shell, $\tau=\rho'\kappa'L'$, which is Lorentz invariant. 
Thus, the radiative energy loss rate in the comoving frame is obtained as follows,
\begin{equation}
\left(\frac{dE'}{dt'}\right)_\mathrm{rad}=\frac{16\pi^2 R'^2}{3\tau}J'=\frac{4\pi R'^2}{3\tau}u_\mathrm{ph}',
\end{equation}
where $u_\mathrm{ph}'=4\pi J'$ is the radiation energy density in the comoving frame. 
This equation also gives the radiative energy loss rate in the laboratory frame. 

Next, we obtain the relation between the zeroth moments $J'$ and $J$ in the comoving and laboratory frames to express the radiative energy loss rate in terms of the radiation energy density in the laboratory frame. 
The Lorentz transformations of the intensity and the direction cosine lead to
\begin{equation}
I=\Gamma_\mathrm{s}^4(1+\beta_\mathrm{s}\mu')^4(J'+\mu'H'),
\label{eq:J+muH}
\end{equation}
and
\begin{equation}
\mu=\frac{\mu'+\beta_\mathrm{s}}{1+\beta_\mathrm{s}\mu'},
\end{equation}
where $\Gamma_\mathrm{s}=(1-\beta_\mathrm{s}^2)^{-1/2}$. 
The zeroth angular moment of this equation leads to
\begin{eqnarray}
J&=&\frac{1}{2}\int_{-1}^{1}\Gamma_\mathrm{s}^4(1+\beta_\mathrm{s}\mu')^4(J'+\mu'H')d\mu
\nonumber\\
&=&\frac{1}{2}\int_{-1}^{1}\Gamma_\mathrm{s}^2(1+\beta_\mathrm{s}\mu')^2(J'+\mu'H')d\mu'
=\Gamma_\mathrm{s}^2\left[\left(1+\frac{\beta_\mathrm{s}^2}{3}\right)J'+\frac{2\beta_\mathrm{s}}{3}H'\right].
\end{eqnarray}
Substituting Equation (\ref{eq:H2}) into the R.H.S. of this equation, one obtains the following relation between $J'$ and $J$,
\begin{equation}
J'=\frac{(1-\beta_\mathrm{s}^2)\tau J}{(1+\beta_\mathrm{s}^2/3)\tau+2\beta_\mathrm{s}/3},
\label{eq:J'}
\end{equation}
which also holds for the radiation energy densities, $u_\mathrm{ph}'$ and $u_\mathrm{ph}$,
\begin{equation}
u_\mathrm{ph}'=\frac{(1-\beta_\mathrm{s}^2)\tau u_\mathrm{ph}}{(1+\beta_\mathrm{s}^2/3)\tau+2\beta_\mathrm{s}/3}.
\end{equation}
Thus, the radiative energy loss rate  can be expressed in terms of the radiation energy density in the laboratory frame as follows,
\begin{equation}
\left(\frac{dE}{dt}\right)_\mathrm{rad}=4\pi R^2u_\mathrm{ph}\frac{1-\beta_\mathrm{s}^2}{(3+\beta_\mathrm{s}^2)\tau+2\beta_\mathrm{s}}.
\label{eq:dEdt_rad}
\end{equation}

\section{Convergence Check of Numerical Simulations}
\begin{figure}[tbp]
\begin{center}
\includegraphics[scale=0.5]{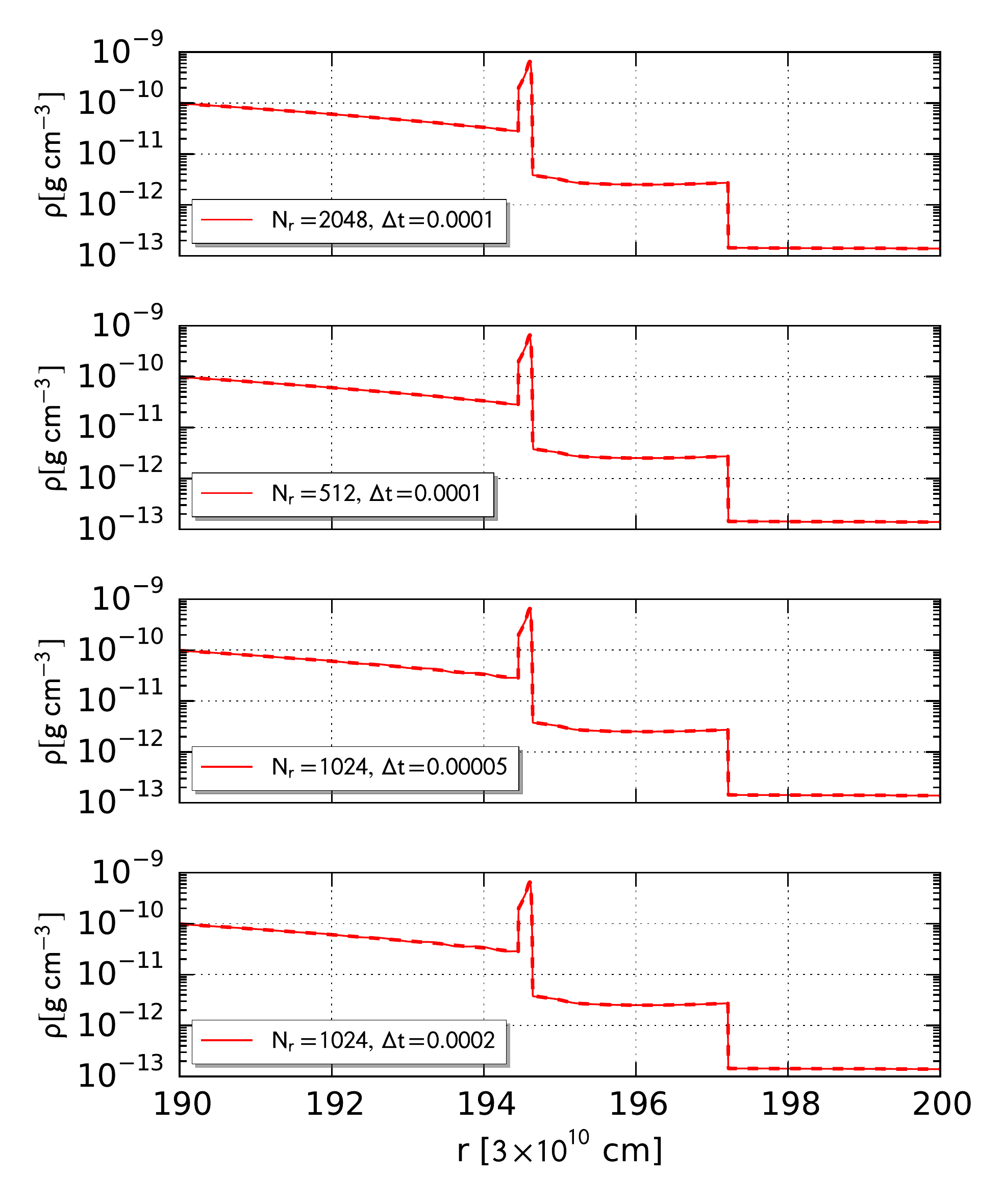}
\caption{Results of the convergence study of the numerical simulations. 
The solid lines show the radial density profiles from the simulations with different spacial and temporal resolutions, while the dashed line in each panel shows that of the fiducial simulation, which is the same model in Figures \ref{fig:evolution} and \ref{fig:profile}.}
\label{fig:resolution}
\end{center}
\end{figure}

In this section, we provide results of the convergence study of our numerical simulations. 
In the simulations performed in Section \ref{sec:numerical}, the whole numerical domain is covered by $N_r=1024$ at the lowest AMR level and the time step is set to $\Delta t=0.0001$. 
We carry out additional four simulations with the same parameter as the model in Figures \ref{fig:evolution} and \ref{fig:profile} but with different spatial and temporal resolutions. 
The number of the numerical cells at the lowest AMR level and the time step are doubled and halved compared to the fiducial model, $(N_r,\Delta t)=(2048,0.0001)$,\ $(512,0.0001)$, $(1024,0.00005)$, and $(1024,0.0002)$. 
The evolution is followed up to $t=200$ s. 
Figure \ref{fig:resolution} shows the results. 
The panels show the radial profiles of the density from $r=190$ to $r=200$ in units of $3\times 10^{10}$ cm at $t=200$ s. 
In each panel, the result of the simulation is plotted as a solid line and compared with the fiducial model (dashed line) with $(N_r,\Delta t)=(1024,0.0001)$. 
The different models show quite similar density profiles, proving the convergence of the simulation shown in this paper.

\end{document}